\documentclass[journal]{IEEEtran}
\usepackage{ifpdf}

\usepackage{amssymb}

\usepackage{cite}

\usepackage[pdftex]{graphicx}
\ifCLASSINFOpdf
  \usepackage[pdftex]{graphicx}
   \DeclareGraphicsExtensions{.pdf,.jpeg,.png}
\usepackage{amsmath}
\usepackage{upgreek}

\usepackage{algorithmic}

\usepackage{array}

\usepackage{booktabs}
\usepackage{multirow}

\ifCLASSOPTIONcompsoc
  \usepackage[caption=false,font=normalsize,labelfont=sf,textfont=sf]{subfig}
\else
  \usepackage[caption=false,font=footnotesize]{subfig}
\fi
\usepackage{fixltx2e}
\usepackage{dblfloatfix}

\ifCLASSOPTIONcaptionsoff
  \usepackage[nomarkers]{endfloat}
 \let\MYoriglatexcaption\caption
 \renewcommand{\caption}[2][\relax]{\MYoriglatexcaption[#2]{#2}}
\fi

\usepackage{amsmath}
\usepackage{bm}
\usepackage{float}

\usepackage{url}

\begin{document}
%

\title{\textit{End-to-end} Learning for Spectrum Monitoring applications in future 5G networks}

\title{\textit{End-to-end} Learning from Spectrum Data: \\
A unified approach for Signal Identification in Spectrum Monitoring applications for 5G}

\title{\textit{End-to-end} Learning from Spectrum Data: \\
A Deep Learning approach for Wireless Signal Identification in Spectrum Monitoring applications}


\author{\IEEEauthorblockN{Merima Kulin\IEEEauthorrefmark{1},
Tarik Kazaz\IEEEauthorrefmark{2},~\IEEEmembership{Student Member,~IEEE},
Ingrid Moerman\IEEEauthorrefmark{1} and Eli de Poorter\IEEEauthorrefmark{1},~\IEEEmembership{Member,~IEEE}} \\
\thanks{Corresponding author: Merima Kulin (e-mail: merima.kulin@ugent.be)}}

%

%



\IEEEtitleabstractindextext{%
\begin{abstract}

This paper presents \textit{end-to-end} learning from spectrum data - an umbrella term for new sophisticated wireless signal identification approaches in spectrum monitoring applications based on deep neural networks.
End-to-end learning allows to (i) automatically learn features directly from simple wireless signal representations, without requiring design of hand-crafted expert features like higher order cyclic moments, and (ii) train wireless signal classifiers in one end-to-end step which eliminates the need for complex multi-stage machine learning processing pipelines.
The purpose of this article is to present the conceptual framework of end-to-end learning for spectrum monitoring and systematically introduce a generic methodology to easily design and implement wireless signal classifiers. Furthermore, we investigate the importance of the choice of wireless data representation to various spectrum monitoring tasks. In particular, two case studies are elaborated (i) modulation recognition and (ii) wireless technology interference detection. For each case study three convolutional neural networks are evaluated for the following wireless signal representations: temporal IQ data, the amplitude/phase representation and the frequency domain representation. 
From our analysis we prove that the wireless data representation impacts the accuracy depending on the specifics and similarities of the wireless signals that need to be differentiated, with different data representations resulting in accuracy variations of up to 29\%.
Experimental results show that using the amplitude/phase representation for recognizing modulation formats can lead to performance improvements up to 2\% and 12\% for medium to high SNR compared to IQ and frequency domain data, respectively. For the task of detecting interference, frequency domain representation outperformed amplitude/phase and IQ data representation up to 20\%.



\end{abstract}

}

\maketitle

\IEEEdisplaynontitleabstractindextext

%
\IEEEpeerreviewmaketitle

\section{Introduction}
\label{sec:intro}
%
%
%
%

\IEEEPARstart{W}{ireless} networks are currently experiencing a dramatic evolution. Some trends observed are the increasing number and diversity of wireless devices, with an increasing spectrum demand.
Unfortunately, the radio frequency spectrum is a scarce resource.
As a result, particular parts of the spectrum are used heavily whereas other parts are vastly underutilized \cite{hoyhtya2016spectrum}. For example, the unlicensed bands are extremely overutilized and suffer from cross-technology interference \cite{hithnawi2014understanding}.



It is indisputable that \textit{monitoring} and \textit{understanding} the spectrum resource usage will become a critical asset for 5G in order to improve and regulate the radio spectrum utilization.
However, monitoring the spectrum use in such a complex wireless system requires distributed sensing over a wide frequency range, resulting in a radio spectrum data deluge \cite{ding2014big}.
Extracting meaningful information about the spectrum usage from \textit{massive} and complex spectrum datasets requires sophisticated and advanced algorithms. 
This paves the way for new innovative spectrum access schemes and the development of novel identification mechanisms that will provide awareness about the radio environment. 
For instance, technology identification, modulation type recognition and interference source detection are essential for interference mitigation strategies to continue effective use of the scarce spectral resources and enable the coexistence of heterogeneous wireless networks.

In this paper, we investigate \textit{end-to-end} learning from spectrum data as a unified approach to tackle various challenges related to the problems of inefficient spectrum management, utilization and regulation that the next generation of wireless networks is facing. 
Whether the goal is to recognize a technology or a particular modulation type, identify the interference source or an interference-free frequency channel, we argue that the various problems may be treated as a generic problem type that we refer to as \textit{wireless signal identification}, which is a natural target for machine learning classification techniques.
In particular, \textit{end-to-end learning} refers to processing architectures where the entire pipeline, connecting the input (i.e the data representation of a sensed wireless signal) to the desired output (i.e. the predicted type of signal), is learned purely from data \cite{muller2006off}. 
This setting simplifies the overall system design and completely eliminates the need for designing expert features such as higher order cyclic moments, and results in accurate wireless signal classifiers.




\subsection{Scope and Contributions}
This paper provides a comprehensive introduction to \textit{end-to-end} learning from spectrum data. 
The main contributions of this paper are as follows:
\begin{itemize}
\item Potential end-to-end learning use cases for spectrum monitoring are identified.
In particular, two categories are presented. The first category are use cases where detecting spectral opportunities and spectrum sharing is necessary such as in cognitive radio and emerging cognitive IoT networks. The second, are scenarios where detecting radio emitters is needed such as in spectrum regulation.
\item To set a preliminary background on this interdisciplinary topic a brief introduction to machine learning/deep learning is provided and their role for spectrum monitoring is discussed. Then, a reference model for deep learning for spectrum monitoring applications is defined.
\item A conceptual framework for end-to-end learning is proposed, followed by a comprehensive overview of the methodology for collecting spectrum data, designing wireless signal representations, forming training data and training deep neural networks for wireless signal classification tasks.
\item To demonstrate the approach, experiments are carried out for two case studies: (i) modulation recognition and (ii) wireless technology interference detection, that demonstrate 
the impact of the choice of wireless data representation on the presented results. 
For modulation recognition, the following modulation techniques are considered: BPSK (binary phase shift keying), QPSK (quadrature phase shift keying), \textit{m}-PSK (phase shift keying, for $m=8$), \textit{m}-QAM (quadrature amplitude modulation, for $m=16$ and $64$), CPFSK (continuous phase frequency shift keying), GFSK (Gaussian frequency shift keying) and \textit{m}-PAM (pulse amplitude modulation for $m=4$). For wireless technology identification, three representative technologies operating in the unlicensed bands are analysed: IEEE 802.11b/g, IEEE 802.15.4 and IEEE 802.15.1.
\end{itemize}

The rest of the paper is organized as follows. The remainder of Section \ref{sec:intro} presents related work. Section \ref{sec:usecases} presents motivating scenarios for the proposed approach. Section \ref{sec:basics} introduces basic concepts related to 
machine learning/deep learning concluded with a high-level processing pipeline for their 
application to spectrum monitoring scenarios.
Section \ref{sec:end-to-end} presents the end-to-end learning methodology for wireless signal classification. In Section \ref{sec:evaluation} the methodology is applied to two scenarios and experimental results are discussed. Section \ref{sec:oc} discusses open challenges related to the implementation and deployment of future end-to-end spectrum management systems. Section \ref{sec:concl} concludes the paper.

\subsection{Related work}

\textbf{Traditional signal identification}. Previous research efforts in wireless communication related to signal identification are dominantly based on signal processing tools for communication \cite{axell2012spectrum} such as cyclostationary feature detection \cite{kim2007cyclostationary}, sometimes in combination with traditional machine learning techniques \cite{fehske2005new} (e.g. support vector machines (SVM), decision trees, k-nearest neighbors (k-NN), neural networks, etc.). The design of these specialized solutions have proven to be time-demanding as they typically rely on manual extraction of expert features for which a significant amount of domain knowledge and engineering is required.

\textbf{Deep learning for signal classification}. Motivated by recent advances and the remarkable success of deep learning, especially convolutional neural networks (CNN), in a broad range of problems such as image recognition, speech recognition and machine translation \cite{krizhevsky2012imagenet},
wireless communication engineers recently used similar approaches to improve on the state of the art in signal identification tasks in wireless networks. 
One of the pioneers in the domain were the authors of
\cite{o2016convolutional}, who demonstrated that CNNs trained on time domain IQ data significantly outperform traditional approaches for automatic modulation recognition based on expert features such as cyclic-moment based features, and conventional classifiers such as 
decision trees, k-NNs, SVMs, NN and Naive Bayes.
Selim et al. \cite{selim2017spectrum} propose to use amplitude and phase difference data to train CNN classifiers able to detect the presence of radar signals with high accuracy.
Akeret at al. \cite{akeret2017radio} propose a novel technique to accurately detect radio frequency interference in radio astronomy by training a CNN on 2D time domain data acquired from a radio telescope.
The authors of \cite{schmidt2017wireless} propose a novel method for interference identification in unlicensed bands using CNNs trained on frequency domain data \cite{schmidt2017wireless}. Several wireless technologies (e.g. DVB, GSM, LTE...) have been classified with high accuracy in \cite{rajendran2017distributed} using deep learning on averaged magnitude FFT data.

These individual works focus on specific deep learning applications pertaining to wireless signal classification using particular data representations. They do not provide a detailed methodology necessary to understand how to apply the same approach to other potential use cases, neither they provide sufficient information as a guide for selecting a wireless data representations.
This information is necessary for someone aiming to reproduce existing attempts, build upon it or to generate new application ideas.

\textbf{Deep learning for wireless networks}. Recently, the authors of \cite{o2017introduction} provided an overview of the state-of-the art and potential future deep learning applications in wireless communication. 
The authors of \cite{yao2017deepsense} propose a unified deep learning framework for mobile sensing data.
However, none of these studies focuses on spectrum monitoring scenarios and the underlying data models for training wireless signal classifiers.

To remedy these shortcomings, this paper presents \textit{end-to-end learning from spectrum data}: a deep learning framework for solving various wireless signal classification problems for spectrum monitoring applications in a unified manner.
To the best of our knowledge, this article is the first comprehensive work that elaborates in detail the methodology for (i) collecting, transforming and representing spectrum data, (ii) designing and implementing data-driven deep learning classifiers for wireless signal identification problems, and that (iii) looks at several data representations for different classification problems at once.
The technical approach depicted in this paper is deeply interdisciplinary and systematic, calling for the synergy of expertise of computer scientists, wireless communication engineers, signal processing and machine learning experts with the ultimate aim of breaking new ground and raising awareness of this emerging interdisciplinary research area.
Finally, this paper is at an opportune time, when (i) recent advances in the field of machine learning, (ii) computational advances and parallelization used to speed up training and (iii) efforts in making large amounts of spectrum data available, have paved the way for novel spectrum monitoring solutions.

\textbf{Notation and terminology.} We indicate a scalar-valued
variable with normal font letters (i.e. $x$ or $X$). Matrices will be denoted using bold capitals such as $\textbf{X}$. Vectors will be denoted with a bold lower case letter (i.e. $\textbf{x}$), which may sometimes appear as row or column vectors of a matrix (i.e. $\textbf{x}_k$ is the $k$-th column vector). With $x_i$ and $x_{ij}$ we will indicate the entries
of $\textbf{x}$ and $\textbf{X}$, respectively. The notation $()^T$ denotes the \textit{transpose} of a matrix or vector, while $()^*$ denotes complex conjugation. 
We indicate by $||\textbf{x}||_p= (\sum_{n=0}^{N-1}|x_n|^p)^{1/p} $ the $l_p$-norm of vector $\textbf{x}$.

\section{Characteristic Use Cases for end-to-end learning from Spectrum Data}
\label{sec:usecases}
End-to-end learning from spectrum data is a new approach that can automatically learn features directly from simple wireless signal representations, without requiring design of hand-crafted expert features like higher order cyclic moments. The term \textit{end-to-end} refers to the fact that the learning procedure can train wireless signal classifiers in one end-to-end step which eliminates the need for complex multi-stage expert machine learning processing pipelines.

\begin{figure}[!t]
    \centering
    \includegraphics[trim={1.2cm 0 0 0}, clip, width=0.5\textwidth]{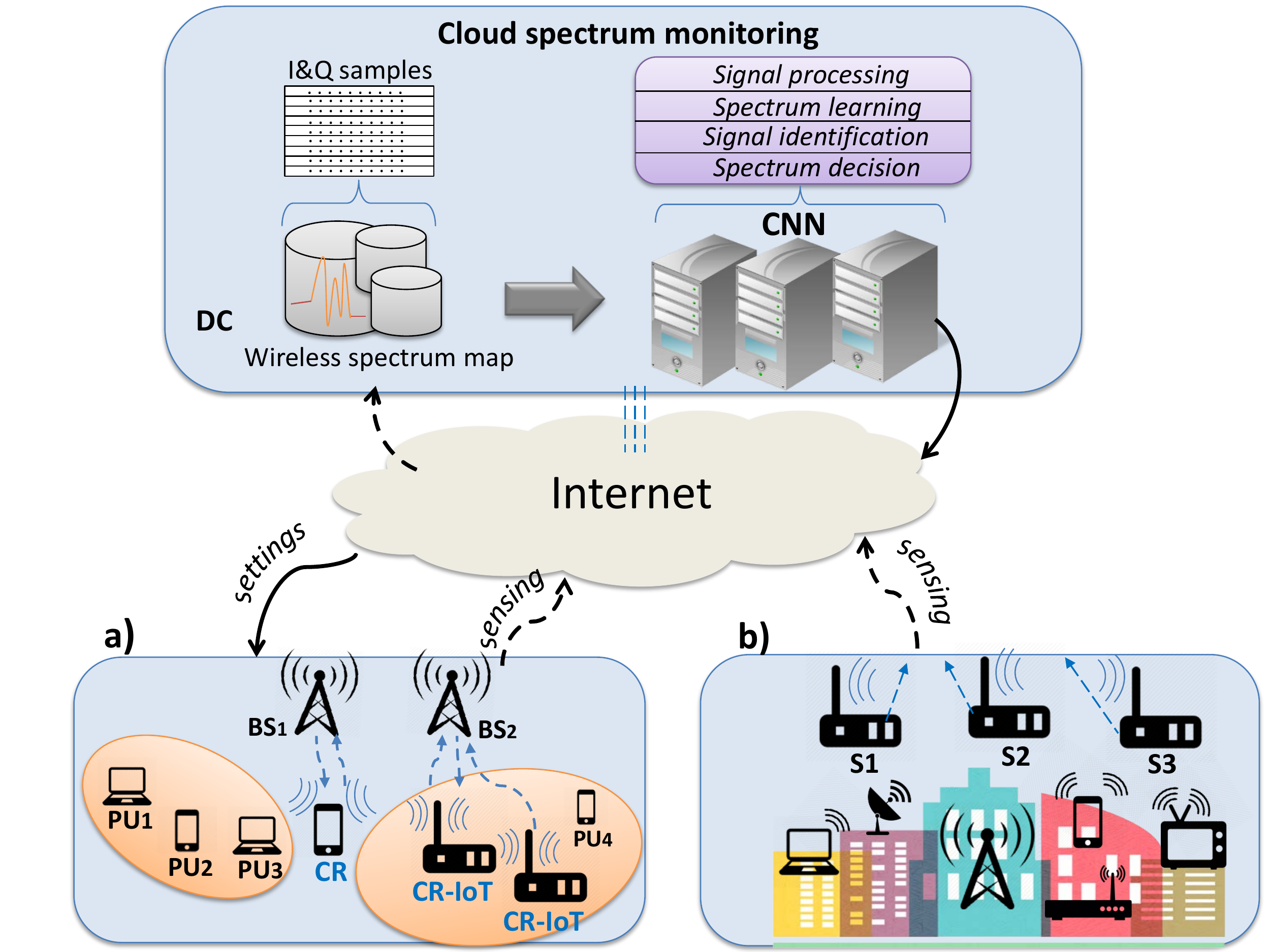}
    \caption{Data-driven CNN-based flexible spectrum management framework}
    \label{fig:architec}
\end{figure}

Before diving deep into the concept of end-to-end learning from spectrum data, we first consider the architecture presented on Figure \ref{fig:architec} with two motivating scenarios that illustrate characteristic use-cases for the presented approach.

\textit{\textbf{Detecting spectral opportunities \& Spectrum Sharing}}

\textit{	1) Cognitive radio}

The ever-increasing radio spectrum demand combined with the currently dominant fixed spectrum policy assignment \cite{chen2009mining}, have inspired the concepts of cognitive radio (CR) and dynamic spectrum access (DSA) aiming to improve  radio spectrum utilization.
A CR network (CRN) is an intelligent wireless communication system that is aware of its radio environment, i.e. spectral opportunities, and can intelligently adapt its operating parameters by interacting and learning from the environment \cite{haykin2005cognitive}. In this way, the CRN can infer the spectrum occupancy to identify unoccupied frequency bands (white spaces/spectrum holes) and share them with licensed users (primary users (PU))
in an opportunistic manner \cite{akyildiz2008survey}.

Figure \ref{fig:architec} a) shows the basic operational process of a data-driven CRN.
First, CR users intermittently sense its surrounding radio environment and report their sensing results via a control channel to a nearby base station (BS). Then, the BS forwards the request to a back-end data center (DC), which combines the crowdsourced sensing information from several CR users into a spectrum map. The DC infers the spectrum use in order to determine the presence of PUs (a characteristic wireless signal) and diffuses the spectrum availability information back to the cognitive users.
For this purpose, the DC first learns a CNN model offline based on the sensing reports, and then employs the model to discriminate between a spectrum hole and an occupied frequency channel.
 
\textit{	2) Cognitive IoT}

The Internet of Things (IoT) paradigm envisioned a world of "always connected" devices/objects/\textit{things} to the Internet \cite{gubbi2013internet}. In this world, heterogeneous wireless technologies and standards emerge operating in the unlicensed frequency bands, which puts enormous pressure on the available spectrum. 
The increasing wireless spectrum demand rises several communication challenges such as co-existence, cross-technology interference and scarcity of interference-free spectrum bands \cite{gollakota2011clearing, hithnawi2014understanding}.
To address these challenges, recent research work proposed a CR-based IoT \cite{khan2017cognitive, wu2014cognitive} to enable dynamic spectrum sharing among heterogeneous wireless networks.

Figure \ref{fig:architec} a) depicts this situation. It can be seen that CR-IoT devices are equipped with cognitive functionalities allowing them to search for interference-free spectrum bands and accordingly \textit{reconfigure} their transmission parameters. First, CR-IoT devices send spectrum sensing reports to a CNN-based DC. Then, the DC learns and estimates the presence of other emitters and uses that information to detect interference sources and interference-free channels. 
This enables smart and effective interference mitigation and spectrum management strategies for co-existence with CR and legacy technologies and modulation types.

\textit{\textbf{Spectrum management policy and regulation}}

Spectrum regulatory bodies continuously monitor the radio frequency spectrum use to prevent users from harmful interference and allow optimum use thereof \cite{staple2004end}.
Interference may be a result of unauthorized emissions, electromagnetic interference (EMI) and devices that operate beyond technical specifications. In order to resolve problems associated with wireless interference, spectrum managers traditionally use a combination of engineering analysis and data obtained from spectrum measurements.
However, in the era of today's "wireless abundance", where various services and wireless technologies share the same frequency bands, the identification of unauthorized transmitters can be very difficult to achieve. More intelligent algorithms are needed that can automatically mine the spectrum data and identify interference sources.

Figure \ref{fig:architec} b) presents a CNN-based spectrum management framework for spectrum regulation.
Deployed sensor devices, e.g. $\{S1, S2, S3\}$, collect spectrum measurements and contribute their observations to a DC to create interference maps. 
The DC uses signal processing  techniques together with a CNN model to mine the obtained spectrum data and identify existing interferers.
The mined patterns  are key for ensuring compliance with national and international spectrum management regulations.

\section{The role of Deep Learning in Spectrum monitoring}
\label{sec:basics}

There are two goals of this section. The first is to introduce the key ideas underlying machine learning/deep learning. The second is to derive a reference model for machine learning/deep learning applications for spectrum monitoring, management and spectrum regulation.

\subsection{Machine Learning}
\label{ml}

Machine learning (ML) refers to a set of algorithms that \textit{learn} a statistical model from historical data. The obtained model is \textit{data-driven} rather then explicitly derived using domain knowledge.

\subsubsection{Preliminaries}
The goal of ML is to find a mathematical function, $f$, that defines the relation between a set of inputs $X$, and a set of outputs $Y$, i.e.
\begin{equation} \label{eq:1}
f: X \rightarrow Y
\end{equation}

The inputs, $\textbf{X} \in \mathcal{R}^{mxn}$, present a number of distinct data points, samples or observations denoted as 
\begin{equation}
\textbf{X}=
\begin{bmatrix}
	{\textbf{x}_1}^T \\
	{\textbf{x}_2}^T \\
	\vdots \\
	{\textbf{x}_m}^T 
\end{bmatrix}
\end{equation}

where $m$ is the sample size, while $\textbf{x}_i \in \mathcal{R}^{n}$ is a vector of $n$ measurements or \textit{features} for the $i$th observation called a \textit{feature vector},
\begin{equation}
\textbf{x}_i=[x_{i1}, x_{i2}, ..., x_{in}]^{T}, i=1,...,m
\end{equation}

The outputs, $\textbf{y} \in \mathcal{R}^{m}$, 
are all the outcomes, labels or target values corresponding to the $m$ inputs $\textbf{x}_i$, denoted by
\begin{equation}
\textbf{y}=	[y_1, y_2, ..., y_m]^{T}
\end{equation}

Then the observed data consists of $m$ input-output pairs, called the \textit{training data} or \textit{training set}, $S$,
\begin{equation} \label{dataset}
S=\{(\textbf{x}_1, y_1),(\textbf{x}_2, y_2), ..., (\textbf{x}_m, y_m)\}
\end{equation}

Each pair $(\textbf{x}_i, y_i)$ is called a \textit{training example} because it is used to \textit{train} or teach the learning algorithm  how to obtain $f$.

In machine learning, $f$ is called the predictor whose task is to \textit{predict} the outcome $y_i$ based on the input values of $\textbf{x}_i$.
There are two classical data models depending on the prediction type, described by:

\[
    f(x)=\left\{
                \begin{array}{ll}
                  regressor \text{: } \text{ if } y \in \mathcal{R}	\\
                  classifier \text{: } \text{ if } y \in \{0,1\}	\\

                \end{array}
              \right.
\]
In short, when the output variable $y$ is \textit{continuous} or quantitative, the learning problem is a \textit{regression} problem. But, if $y$ predicts a discrete or \textit{categorical} value, it is a \textit{classification} problem.

\subsubsection{Learning the model}
\label{sec:learnmodel}
Given a training set, $S$, the goal of a machine learning algorithm is to \textit{learn} the mathematical model for $f$.
To make sense of this task, we assume there exists a fixed but unknown distribution, $p(x,y)=p_X(x)p(y|x)$, according to which the data sample is identically and independently distributed (i.i.d). Here, $p_X(x)$ is the marginal distribution that models the uncertainty in the sampling of the input points, while $p(y|x)$ is the conditional distribution that describes the statistical relation between the input and output.

Thus, $f$ is some fixed but unknown function that defines the relation between $X$ and $Y$.
The depicted ML algorithm determines the functional form or shape. 
The unknown function $f$ is estimated by applying the selected learning method to the training data, $S$, so that $f$ is a good estimator for new unseen data, i.e.
\begin{equation}
y \approx \hat{y}=\hat{f}(x_{new})
\end{equation}

The predictor $f$ is parametrized by a vector $\boldsymbol{\uptheta} \in \mathcal{R}^{n} $, and describes a \textit{parametric} model.
In this setup, the problem of estimating $f$ reduces down to one of estimating the parameters $\boldsymbol{\uptheta}=[\theta_1, \theta_2,...,\theta_n]^{T}$.
In most practical applications, the observed data are corrupted versions of the expected values that would be obtained under ideal circumstances. These unavoidable corruptions, typically termed \textit{noise}, prevent the extraction of true parameters from the observations.
With this in regard, the generic data model may be expressed as
\begin{equation}
y=f(\textbf{x})+\boldsymbol{\upepsilon}
\end{equation}
where $f(\textbf{x})$  is the model and $\boldsymbol{\upepsilon}$ are additive measurement errors and other discrepancies.
The goal of ML  is to find the input-output relation that will "best" match the noisy observations.
Hence, the vector $\boldsymbol{\uptheta}$ may be estimated by solving a \textit{\textbf{(convex) optimization}} problem. First, a \textit{loss} or \textit{\textbf{cost function}} $l(\textbf{x}, \textbf{y}, \boldsymbol{\uptheta})$ is set, which is a (point-wise) measure of the error 
between the observed data point $y_i$ and the model prediction $\hat{f}(\textbf{x}_i)$ for each value of $\boldsymbol{\uptheta}$. However, $\boldsymbol{\uptheta}$ is estimated on the whole training data, $S$, not just one example.
For this task, the average loss over all training examples called \textit{training loss}, $J$,  is calculated:
\begin{equation}
J(\boldsymbol{\uptheta}) \equiv J(S, \boldsymbol{\uptheta})=\frac{1}{m}\sum_{(\textbf{x}_i,y_i) \in  S}l(\textbf{x}_i,y_i, \boldsymbol{\uptheta})
\end{equation}
where $S$ indicates that the error is calculated on the instances from the training set and $i=1,...,m$.
The vector $\boldsymbol{\uptheta}$ that minimizes the training loss $J(\boldsymbol{\uptheta})$, that is
\begin{equation}
\operatornamewithlimits{argmin}\limits_{\boldsymbol{\uptheta} \in \mathcal{R}^n}{J(\boldsymbol{\uptheta})}
\end{equation}
will give the desired model. Once the model is estimated, for any given input $\textbf{x}$, the prediction for $y$ can be made with $\hat{y}=\boldsymbol{\uptheta}^T\textbf{x}$.

In engineering parlance, the process of estimating the parameters of a model that is a mapping between input and output observations is called \textbf{\textit{system identification}}. System identification or ML classification techniques are well suited for wireless signal identification problems.

\subsection{Deep Learning}
The prediction accuracy of ML models heavily depends on the choice of the data representation or features used for training. For that reason, much effort in designing ML models goes into the composition of pre-processing and data transformation chains that result in a representation of the data that can support effective ML predictions. Informally, this is referred to as \textit{feature engineering}.
\textit{Feature engineering} is the process of extracting, combining and manipulating features by taking advantage of human ingenuity and prior expert knowledge to arrive at more representative ones, that is 
\begin{equation} 
\phi(\textbf{d}): \textbf{d} \rightarrow \textbf{x}
\end{equation}
i.e. the feature extractor $\phi$ transforms the data vector $\textbf{d} \in \mathcal{R}^d$  into a new form, $\textbf{x} \in \mathcal{R}^n$, more suitable for making predictions. 
The importance of feature engineering highlights the bottleneck of machine learning algorithms: their inability to automatically extract the discriminative information from data.

\textit{\textbf{Feature learning}} is a branch of machine learning that moves the concept of learning from "learning the model" to "learning the features". One popular feature learning method is \textit{deep learning}. In particular, this paper focuses on \textit{convolutional neural networks} (CNN).

\textit{\textbf{Convolutional neural networks}} perform feature learning via non-linear transformations implemented as a series of nested \textit{layers}. 
The input data is a multidimensional data array, called \textit{tensor}, that is presented at the \textit{visible layer}.
This is typically a grid-like topological structure, e.g. time-series data, which can be seen as a 1D grid taking samples at regular time intervals, pixels in images with a 2D layout, a 3D structure of videos, etc. 
Then a series of \textit{hidden layers} extract several abstract features. Those layers are "hidden" because their values are not given. Instead, the deep learning model must determine which data representations are useful for explaining the relationships in the observed data.
Each layer consists of several \textit{kernels} that perform a \textit{convolution} over the input; therefore, they are also referred to as \textit{convolutional layers}.
Kernels are feature detectors, that convolve over the input and produce a transformed version of the data at the output. 
Those are banks of finite impulse response \textit{filters} as seen in signal processing, just learned on a hierarchy of layers.
The filters are usually  multidimensional arrays of parameters that are learnt by the learning algorithm \cite{Goodfellow-et-al-2016} through a  training process called \textit{backpropagation}.

For instance, given a two-dimensional input $x$, a two-dimensional kernel $h$ computes the 2D convolution by
\begin{equation} \label{eq:spatconv}
(x*h)_{i,j}=x[i,j]*h[i,j]=\sum_{n}\sum_{m} x[n,m] \cdot h[i-n][j-m]
\end{equation}
i.e. the dot product between their weights and a small region they are connected to in the input. 

After the convolution, a bias term is added and a point-wise nonlinearity $g$ is applied, forming a \textit{feature map} at the filter output. If we denote the $l$-th feature map at a given convolutional layer as $\textbf{h}^l$, whose filters are determined by the coefficients or \textit{weights} $\textbf{W}^l$, the input $\textbf{x}$ and the bias $b_l$, then the feature map $h^l$ is obtained as follows
\begin{equation} \label{eq:featuremap}
{h^l}_{i,j}=g({(W^l*x)}_{ij}+{b}_l)
\end{equation}
where $*$ is the 2D convolution defined by Equation \ref{eq:spatconv}, while $g(\cdot)$ is the activation function. 
Typically, the \textit{rectifier} activation function is used for CNNs, which is defined by $g(x)=max(0,x)$. Kernels using the rectifier are called \textit{ReLU} (Rectified Linear Unit) and have shown to greatly accelerate the convergence during the training process compared to other activation functions. Others common activation functions are the hyperbolic tangent function (\textit{tanh}), $g(x)=\frac{2}{1+e^{-2x}}-1$, and the \textit{sigmoid} activation $g(x)=\frac{1}{1+e^{-x}}$.



In order to form a richer representation of the input signal, commonly, multiple filters are stacked so that each hidden layer consists of multiple feature maps, $\{h^{(l)}, l = 0,...,L\}$ (e.g., $L=64,128,...$, etc).
The number of filters per layer is a tunable parameter or \textit{hyper-parameter}. Other tunable parameters are the filter size, the number of layers, etc. The selection of values for hyper-parameters may be quite difficult, and finding it commonly is much an art as it is science. An optimal choice may only be feasible by trial and error.
The filter sizes are selected according to the input data size so as to have the right level of  “granularity” that can create abstractions at the proper scale.
For instance, for a 2D square matrix input, such as spectrograms, common choices are $3x3$, $5x5$, $9x9$, etc. For a \textit{wide} matrix, such as a real-valued representation of the complex I and Q samples of the wireless signal in $\mathcal{R}^{2xN}$, suitable filter sizes may be $1x3$, $2x3$, $2x5$, etc.

\begin{figure*}[ht]
    \centering
    \includegraphics[width=0.9\textwidth]{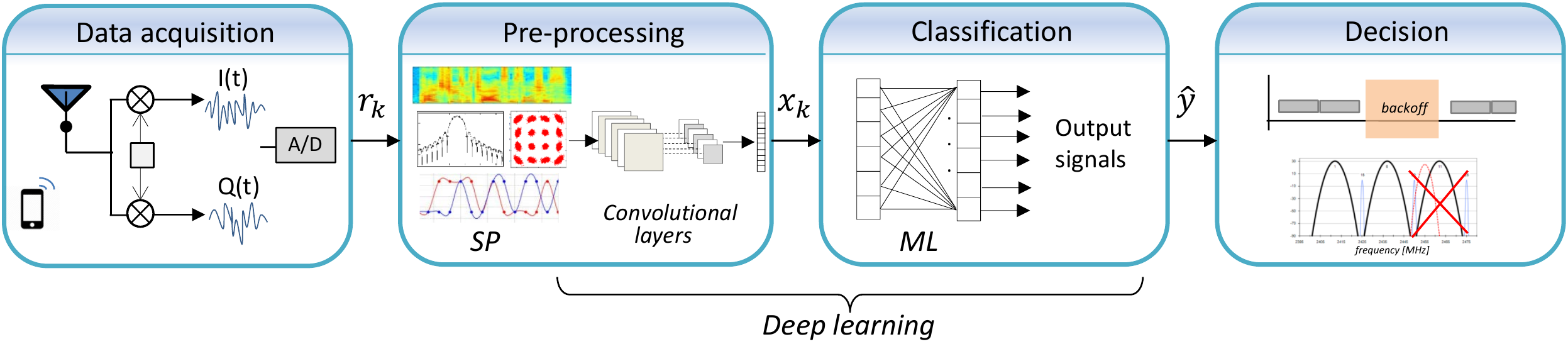}
    \caption{Processing pipeline for end-to-end learning from spectrum data}
    \label{fig:sp_ml}
\end{figure*}

The penultimate layer in a CNN consists of \textit{neurons} that are fully-connected with all feature maps in the preceding layer. Therefore, these layers are called \textit{fully-connected} or \textit{dense} layers.
The very last layer is a \textit{softmax} classifier, which computes the \textit{posterior} probability of each class label over $K$ classes as 
\begin{equation} \label{eq:softmax}
\hat{y_i}=\frac{e^{z_i}}{\sum_{j=1}^{K}e^{z_j}}, \text{  } i=1,...,K
\end{equation}
That is, the scores $z_i$ computed at the output layer, also called \textit{logits}, are translated into probabilities.
A loss function, $l$, is calculated on the last fully-connected layer
that measures the difference between the estimated probabilities, $\hat{y_i}$, and the one-hot encoding of the true class labels, $y_i$.
The CNN parameters, $\boldsymbol{\uptheta}$, are obtained by minimizing the loss function on the training set $\{ x_i, y_i\}_{i \in S}$ of size $m$,
\begin{equation}
\operatornamewithlimits{min}\limits_{\boldsymbol{\uptheta}} \sum_{i \in  S}l(\hat{y_i}, y_i)
\end{equation}
where $l(.)$ is typically the  mean squared error $l(y,\hat{y})=\| y-\hat{y} \|_2^2$ or the \textit{categorical cross-entropy} 
$l(y,\hat{y})={\sum_{i=1}^{m} y_ilog(\hat{y_i})}$ for which a minus sign is often added in front to get the negative \textit{log-likelihood}.


To control over-fitting, typically regularization is used in combination with \textit{dropout}, which is a new extremely effective technique that "drops out" a random set of activations in a layer. Each unit is retained with a fixed probability $p$,
typically chosen using a validation set, or set to $0.5$ which
has shown to be close to optimal for a wide range of applications \cite{srivastava2014dropout}.

\subsection{Deep Learning from spectrum data}
Intelligence capabilities will be of paramount importance in the development of future wireless communication systems to allow them \textit{observe}, \textit{learn} and \textit{respond} to its complex and dynamic operating environment. 
Figure \ref{fig:sp_ml} shows a processing pipeline for realizing intelligent behaviour using deep learning in an end-to-end learning from spectrum data setup. The pipeline consists of:

\textbf{Data acquisition.}
Data is a key asset in the design of future intelligent wireless networks \cite{kulin2016data}.
In order to obtain spectrum data, the radio first \textit{senses} its environment by collecting raw data from various spectrum bands. The raw data consist of $n$ samples, stacked into data vectors $\textbf{r}_k$ which represent the complex envelope of the received wireless signal. These data vectors are the input for end-to-end learning to obtain models that can reason about the presence of wireless signals.

\textbf{Data pre-processing.}
Data pre-processing is concerned with the analysis and manipulation of the collected spectrum data with the aim to arrive at potentially good wireless data representations. 
The raw samples organized into data vectors $\textbf{r}_k$ in the previous block are pipelined as input for signal processing (SP) tools that \textit{analyze}, \textit{process} and \textit{transform} the data to arrive at simple data representations such as frequency, amplitude, phase and spectrum, or more complex \textit{features} $\textbf{x}_k$ such as e.g. cyclostationary features. 
In addition, feature learning such as deep learning may be utilized to automatically extract more low-level and high-level features. 
In many ML applications the choice of features is just as important, if not more important then the choice of the ML algorithm. 

\textbf{Classification.}
The "Classification" processing block enables \textit{intelligence} capabilities to asses the environmental radio context by detecting the presence of wireless signals. This may be the type of the emitters that are utilizing the spectrum (spectrum access scheme, modulation format, wireless technology, etc.), type of interference, detecting an available spectrum band, etc.
We refer to this process as \textit{spectrum learning} \cite{jiang2017machine}.
In future wireless networks ML algorithms may play a key role in automatically \textit{classifying} wireless signals as a step towards intelligent spectrum access and management schemes.

\textbf{Decision.}
The predictions calculated by the ML model are used as input for the decision module. In a CR application, a decision may be related to the best transmission strategy (e.g. frequency band or transmission power) that will maximize the data rate without causing interference to other users. 
This process is called \textit{spectrum decision} \cite{akyildiz2008survey}. 
In the context of CR-IoTs, the decision may relate to an interference mitigation strategy such as back-off for a certain time period.
In other communication scenarios such as spectrum regulation, the decision may relate to a \textit{spectrum policy} or spectrum compliance enforcement applied to a detected source of harmful interference (e.g. fake GSM tower, rouge access point, etc.).

\section{Data-driven end-to-end learning for wireless signal classification}
\label{sec:end-to-end}

The next generation (5G) wireless networks are expected to learn the diverse characteristics of the dynamically changing wireless environment and fluctuating nature of the available spectrum, so as to autonomously determine the optimal system configuration or to support spectrum regulation.

This section introduces a data-driven end-to-end learning framework for spectrum monitoring applications in future 5G networks.
First, the representation of wireless signals used in digital communication and a data model for wireless signal acquisition is introduced. Then, a data model for extracting features, creating training data and designing wireless signal classifiers is presented.
In particular, 
deep learning is used for extracting low-level and higher level wireless signal features and for wireless signal classification.

\subsection{Wireless signal model}
\label{radiosig}

A wireless communication system transmits \textit{information} from one point to another though a wireless medium which is called a \textit{channel}. 
At the system level, a wireless communication model consists of the following parts:

\begin{figure*}[!h]
    \centering
    \includegraphics[width=0.9\textwidth]{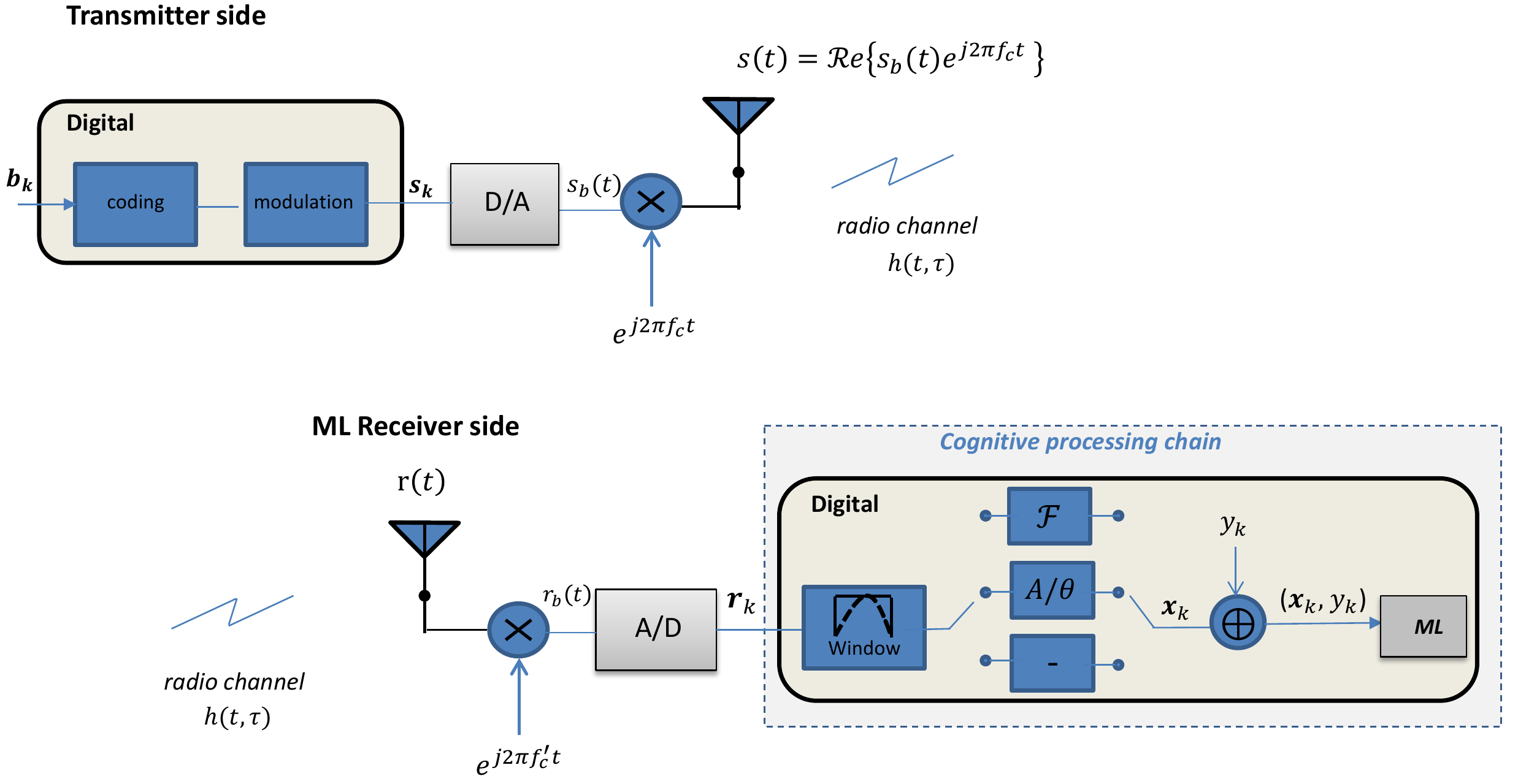}
    \caption{End-to-end learning processing chain to obtain radio spectrum feature vectors}
    \label{fig:dataaq}
\end{figure*}

\textbf{Transmitter}.
The transmitter transforms the message, i.e. a stream of bits, produced by the source of information into an appropriate form for transmission over the wireless channel. Figure \ref{fig:dataaq} shows the processing chain at the transmitter side. First, the bits ${b_k \in \{0,1\}}$ are mapped into a new binary sequence by a \textit{coding} technique. The resulting sequence is mapped to symbols ${s_k}$ from an alphabet or \textit{constellation} which might be real or complex. This process is called \textit{modulation}. 

In the modulation step, the created symbols are mapped to a discrete waveform or signal via a pulse shaping filter and sent to the digital to analog converter module (D/A) where the waveform is transformed into an analog continuous time signal, $s_b(t)$.
The resulting signal is a \textit{baseband} signal that is frequency shifted by the carrier frequency $f_c$ to produce the wireless signal $s(t)$ that is defined by
\begin{multline} 
s(t)=\Re\{s_b(t)e^{j2 \pi f_ct}\}= \\
\Re\{s_b(t)\}\cos(2\pi f_ct)-\Im\{s_b(t)\}\sin(2\pi f_ct)
\end{multline}

where $s(t)$ is a real-valued \textit{bandpass} signal with center frequency $f_c$, while
$s_b(t)=\Re\{s_b(t)\}+j\Im\{s_b(t)\}$ is the baseband complex \textit{envelope} of $s(t)$.

\textbf{Wireless channel}. 
The wireless channel is characterised by the variations
of the channel strength over time and over frequency.
The variations are modeled as (i) \textit{large-scale fading}, which characterizes the path loss of the channel as a function of distance and shadowing by large objects such as buildings and hills, and (ii) \textit{small-scale fading}, which models constructive and destructive interference of the multiple propagation paths between the transmitter and receiver.
The channel effects can be modeled as a linear time-varying system described by a complex finite impulse response (FIR) filter $h(t, \tau)$. If $r(t)$ is the  signal at the channel output, the input/output relation is given by:
\begin{equation} 
r(t)=s(t)*h(t,\tau)
\end{equation}
where $h(t,\tau)$ is the band-limited bandpass channel impulse response, while $*$ denotes the convolution operation.

\textbf{Receiver}. 
The wireless signal at the receiver output will be a corrupted version of the transmitted signal due to 
channel impairments and hardware imperfections of the transmitter and receiver.
Typical hardware related impairments are:
\begin{itemize}
\item \textbf{\textit{Noise}} caused by the resistive components such as the receiver antenna. This thermal noise may be modelled as additive white Gaussian noise (AWGN), $\textbf{n}\sim \mathcal{N}(0,\sigma^2)$.
\item \textbf{\textit{Frequency offset}} caused by the  slightly different local oscillator (LO) signal frequencies at the transmitter, $f_c$, and receiver, ${f_c}'$.

\item \textbf{\textit{Phase Noise}}, $\varphi(t)$, caused by the frequency drift in the LOs used to demodulate the received wireless signal. It causes the angle of the LO signals to drift around its intended instantaneous phase $2\pi f_ct$. 

\item \textbf{\textit{Timing drift}} caused by the difference in sample rates at the receiver and transmitter. 
\end{itemize}

The received wireless signal model can be given by $r(t)=\Re\{r_b(t)e^{j2 \pi f_ct}\}$, where $r_b(t)$ is the baseband complex enveloped defined by 
\begin{equation}
r_b(t)=(s_b(t)* h_b(t,\tau))\dfrac{1}{2} e^{j2 \pi(f_c-{f_c}')t} +n(t)
\end{equation}

where $h_b(t,\tau)$ is the baseband channel equivalent given by
\begin{equation}
h_b(t,\tau)=\sum_{i=0}^{l} \alpha_i(t,\tau) e^{j2 \pi f_c \tau_i(t)+ \varphi_i(t,\tau)}\delta(\tau-\tau_i(t))
\end{equation}

\subsection{Data acquisition}
\label{sec:dataAq}
To derive a machine learning model for wireless signal identification, adequate training data needs to be collected.

Figure \ref{fig:dataaq} summarizes the data acquisition process 
for collecting wireless signal features.
The received signal, $r(t)$, is first amplified, mixed, low-pass filtered and then sent to the analog to digital (A/D) converter, which samples the continuous-time signal at a rate $f_s=1/T_s$ samples per second and generates the discrete version $r_n$.
The discrete signal $r_n=r[nT_S]$ consists of two components, the in-phase, $r_I$, and quadrature component, $r_Q$, i.e.
\begin{equation} 
r_n:=r[n]=r_I[n]+jr_Q[n]
\end{equation}
Suppose, we sample for a period $T$ and collect a batch of $N$ samples. The signal samples $r[n] \in \mathcal{C}$ , $n=0,...,N-1$, are a time-series of complex \textit{raw} samples which may be represented as a data vector. The $k$-th data vector can be denoted as
\begin{equation} \label{eq:complex}
\textbf{r}_k=[r[0],..., r[N-1]]^T
\end{equation}

These data vectors $\textbf{r}_k$ are windowed or segmented representations of the received continuous sample stream, similarly as is seen in audio signal processing. They carry information 
for assessing which type of wireless signal is sensed. This may be the type of modulation, the type of wireless technology, interferer, etc.

\subsection{Wireless signal representation}
\label{sec:sigTransf}
After collecting the $k$-th data vector the ML receiver baseband processing chain transforms it into a new representation suitable for training.  
That is, the $k$-th data vector $\textbf{r}_k \in \mathcal{C}^{N}$ is translated into the $k$-th feature vector $\textbf{x}_k \in \mathcal{R}^{N}$
\begin{equation}
 \textbf{r}_k  \mapsto \textbf{x}_{k}
\end{equation}

This paper considers three simple data representations. 
The first, is a real-valued equivalent of the raw complex temporal wireless signal inspired by the results in \cite{o2016convolutional}. The second, is based on the amplitude and phase of the raw wireless signal, similar to the one used in the work of Selim et al. \cite{selim2017spectrum} for identifying radar signals. The last is a frequency domain representation inspired by the work of Danev et al. \cite{danev2009transient} which showed that frequency-based features outperform their time-based equivalents for  wireless device identification. Each data representation snapshot has a fixed length of $N$ data points.

For each transformation data is visualized to form some intuition about which data representation may provide the most discriminative features for machine learning.
The following data/signal transformations are used:

\textbf{Transformation 1 ($\textbf{IQ}$ vector)}: \textit{The \textbf{IQ vector} is a mapping of the raw complex samples, i.e. data vector $\textbf{r}_k \in \mathcal{C}^{N}$,
into two sets of real-valued data vectors, one that carries the in-phase samples $\textbf{x}_i$ and one that holds the quadrature component values $\textbf{x}_q$. That is}
\begin{equation}
\textbf{x}^{IQ}_k=
\begin{bmatrix}
	{\textbf{x}_i}^T \\
	{\textbf{x}_q}^T 
\end{bmatrix}
\end{equation}
so that $\textbf{x}^{IQ}_k \in \mathcal{R}^{2xN} $.
Mathematically, this may be written as
\begin{align*} 
f: \mathbb{C}^N &\to \mathbb{R}^{2xN} \\
 \textbf{r}_k  &\mapsto \textbf{x}^{IQ}_k
\end{align*}

\textbf{Transformation 2 ($\textbf{A}/\boldsymbol{\upphi}$ vector)}:  \textit{The \textbf{$\textbf{A}/\boldsymbol{\upphi}$ vector} is a mapping from the raw complex data vector $\textbf{r}_k \in \mathcal{C}^{N}$ into two real-valued vectors, one that represents its phase, $\boldsymbol{\upphi}$, and one that represents its magnitude $\textbf{A}$}, i.e.
\begin{equation}
\textbf{x}^{\textbf{A}/\boldsymbol{\upphi}}_k=
\begin{bmatrix}
	{\textbf{x}_A}^T \\
	{\textbf{x}_{\phi}}^T 
\end{bmatrix}
\end{equation}
where $\textbf{x}^{\textbf{A}/\boldsymbol{\upphi}}_k \in \mathcal{R}^{2xN} $, and the \textit{phase}, $\textbf{x}_{\boldsymbol{\upphi}} \in \mathcal{R}^{N}$, and \textit{magnitude} vectors, $\textbf{x}_{A} \in \mathcal{R}^{N}$, have the elements
\begin{equation}
{x_{\phi}}_n=\arctan(\frac{{r_q}_n}{{r_i}_n}) \text{ ,   }
{x_A}_n=({r_q}_n^2 + {r_i}_n^2)^{1/2}\text{, }n=0,...,N-1
\end{equation}

In short, this may be written as
\begin{align*} 
f: \mathbb{C}^N &\to \mathbb{R}^{2xN} \\
 \textbf{r}_k  &\mapsto \textbf{x}^{\textbf{A}/\boldsymbol{\upphi}}_k
\end{align*}

\begin{figure}[!h]
\centering
\subfloat[\textbf{BPSK}]{\includegraphics[trim={0 0 0 1.4cm},clip,width=0.19\textwidth]{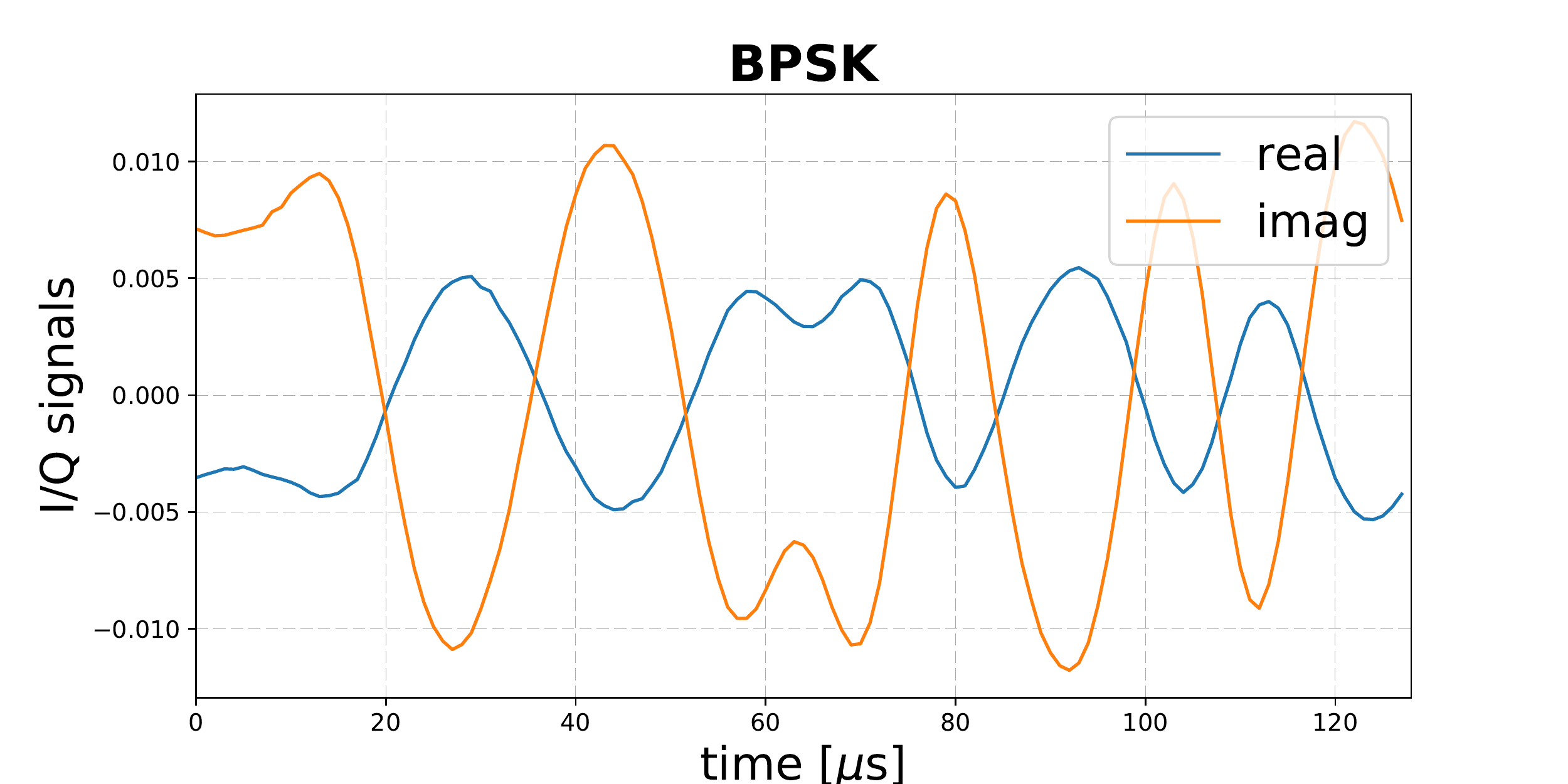}}
\subfloat[\textbf{QPSK}]{\includegraphics[trim={0 0 0 1.4cm},clip,width=0.19\textwidth]{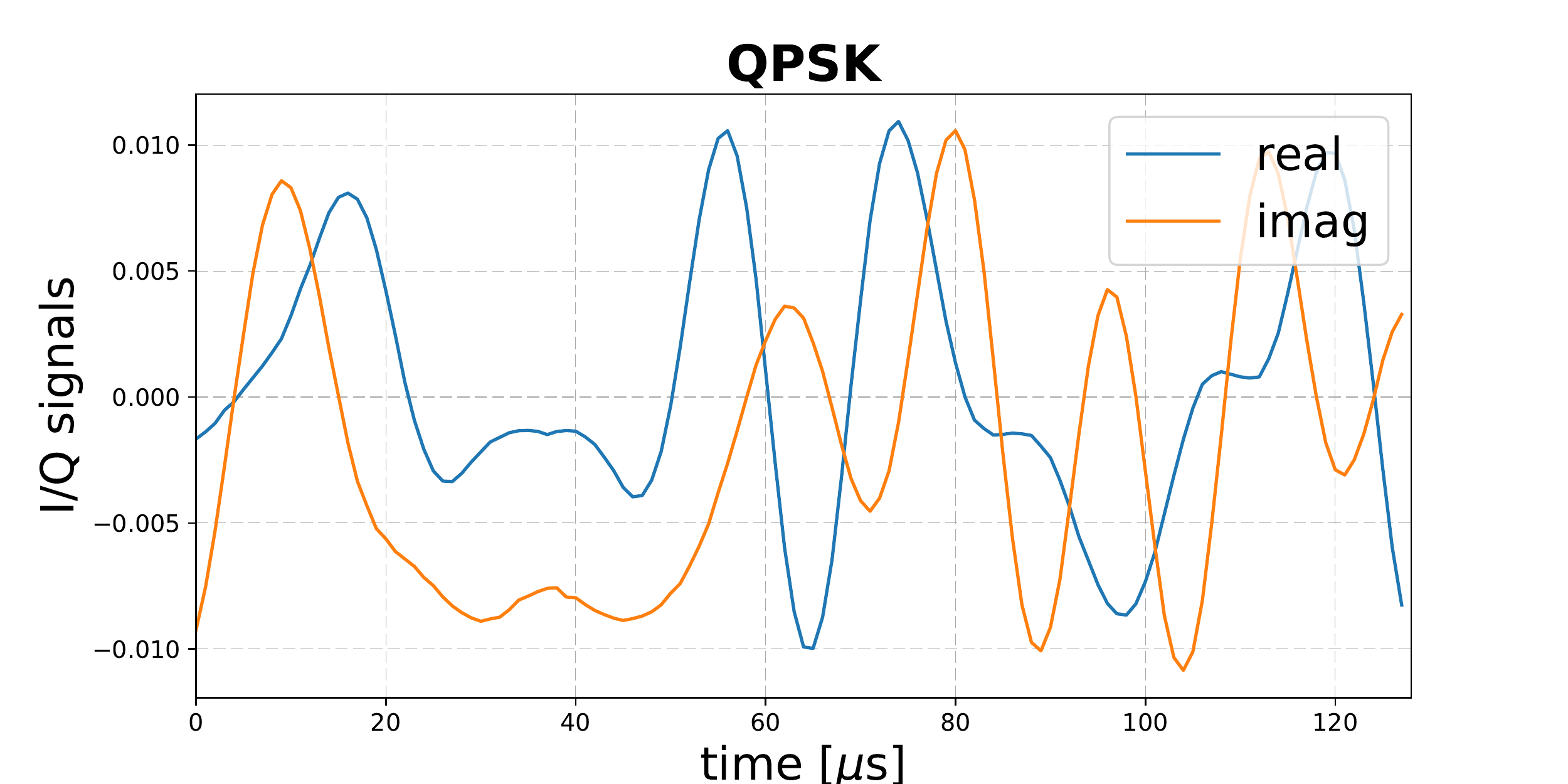}}
\hfil
\subfloat[\textbf{8PSK}]{\includegraphics[trim={0 0 0 1.4cm},clip,width=0.19\textwidth]{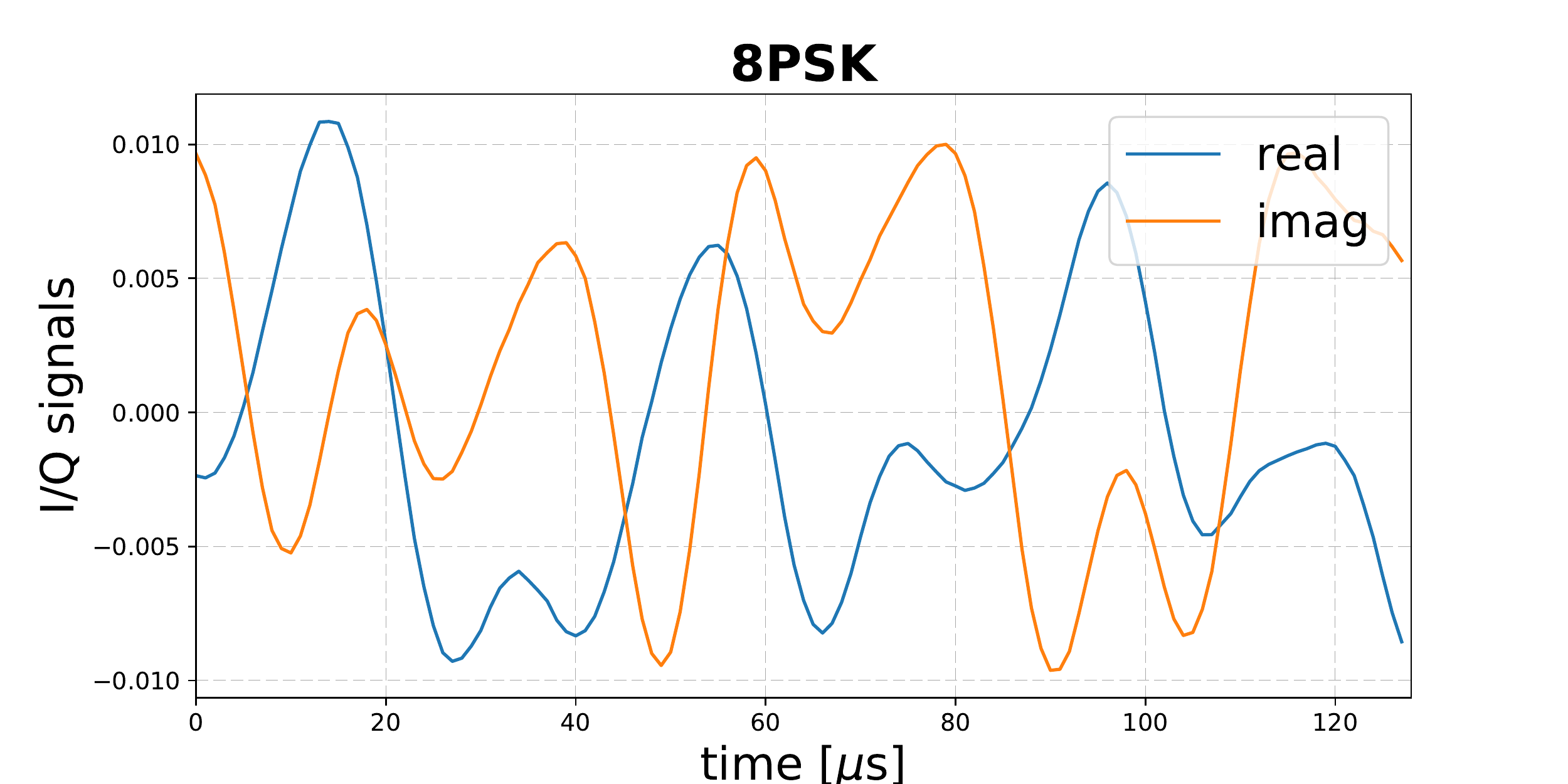}}
\subfloat[\textbf{QAM16}]{\includegraphics[trim={0 0 0 1.4cm},clip,width=0.19\textwidth]{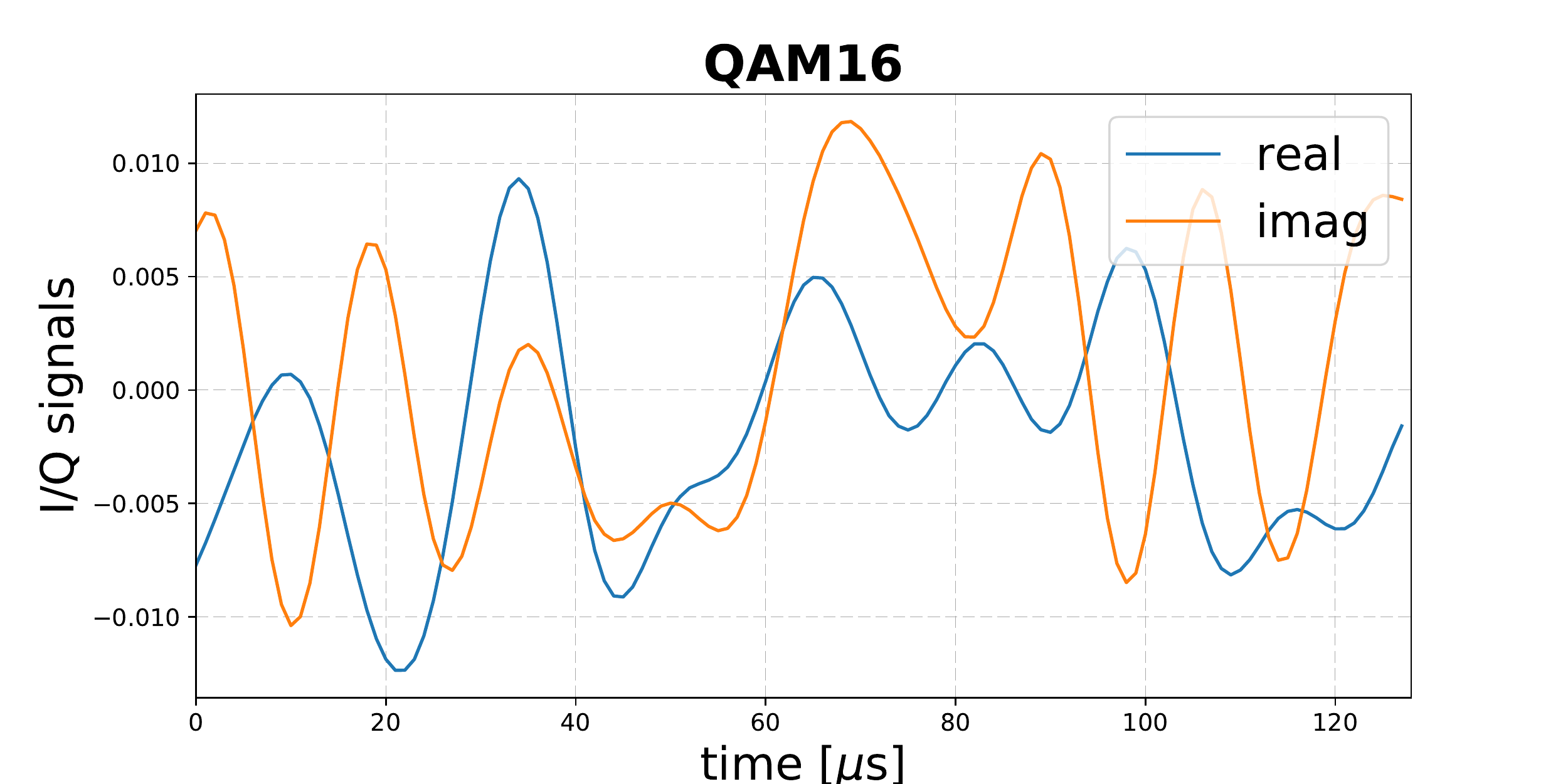}}
\hfill
\subfloat[\textbf{QAM64}]{\includegraphics[trim={0 0 0 1.4cm},clip,width=0.19\textwidth]{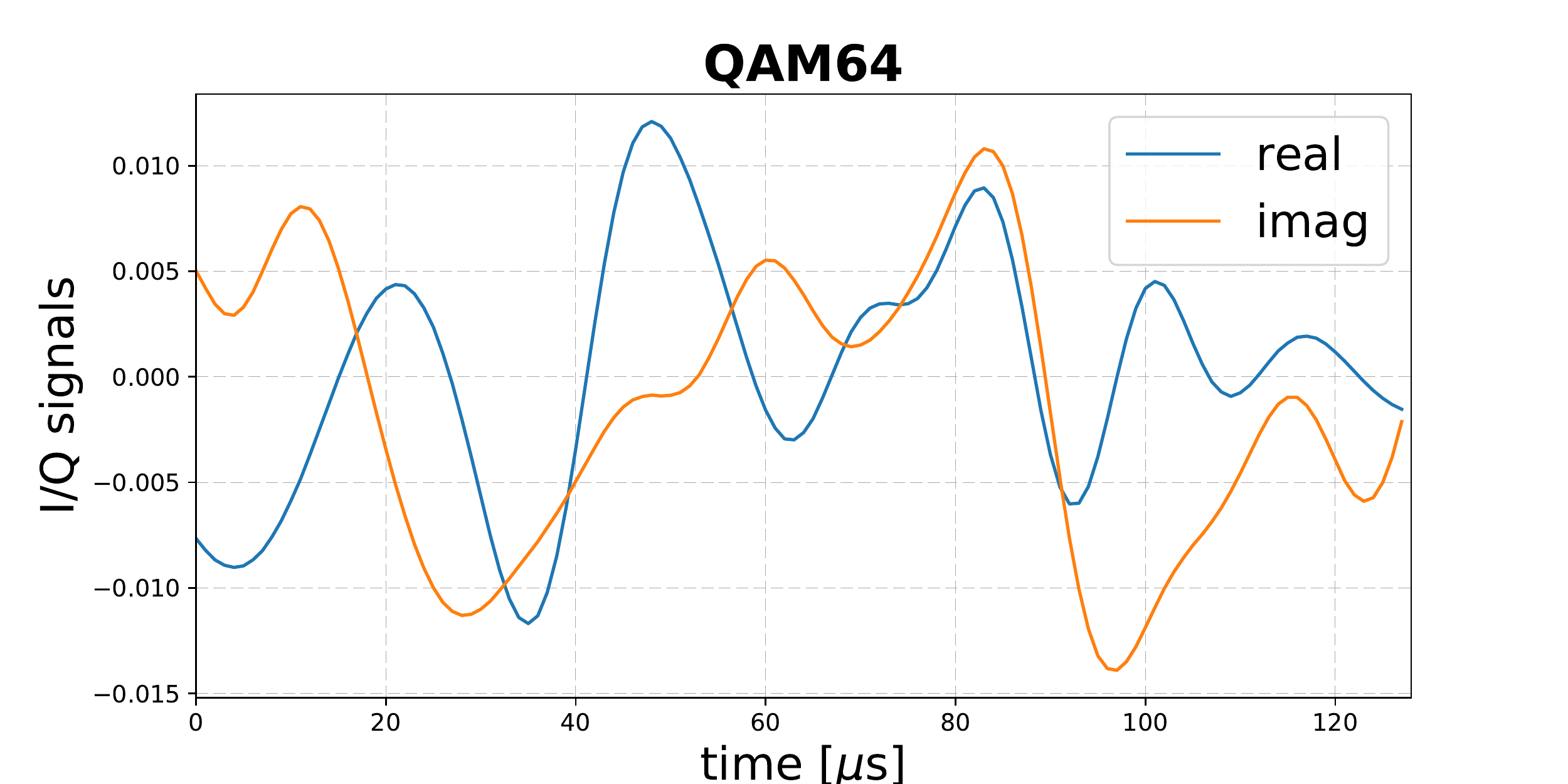}}
\subfloat[\textbf{CPFSK}]{\includegraphics[trim={0 0 0 1.4cm},clip,width=0.19\textwidth]{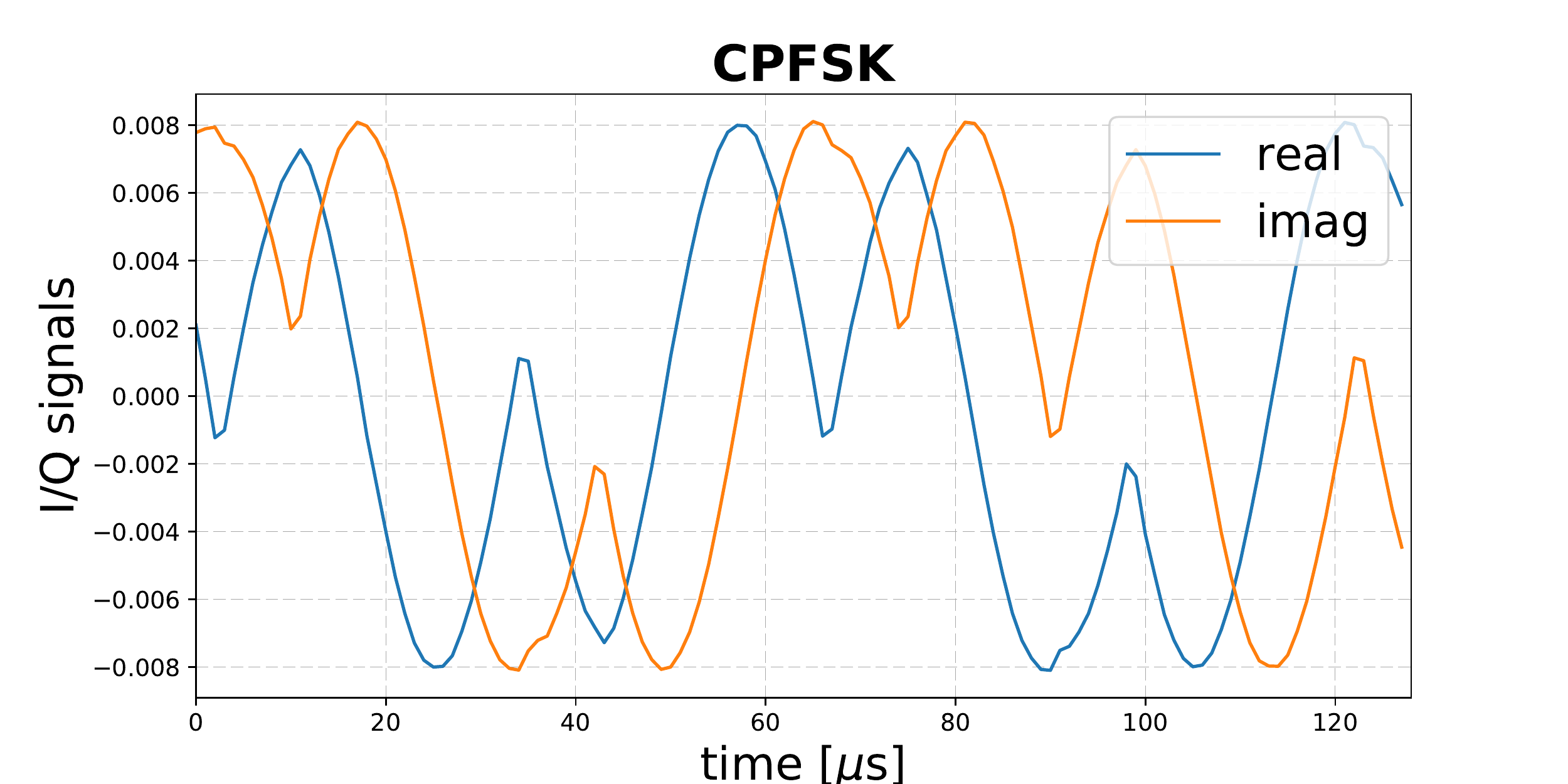}}
\hfill
\subfloat[\textbf{GFSK}]{\includegraphics[trim={0 0 0 1.4cm},clip,width=0.19\textwidth]{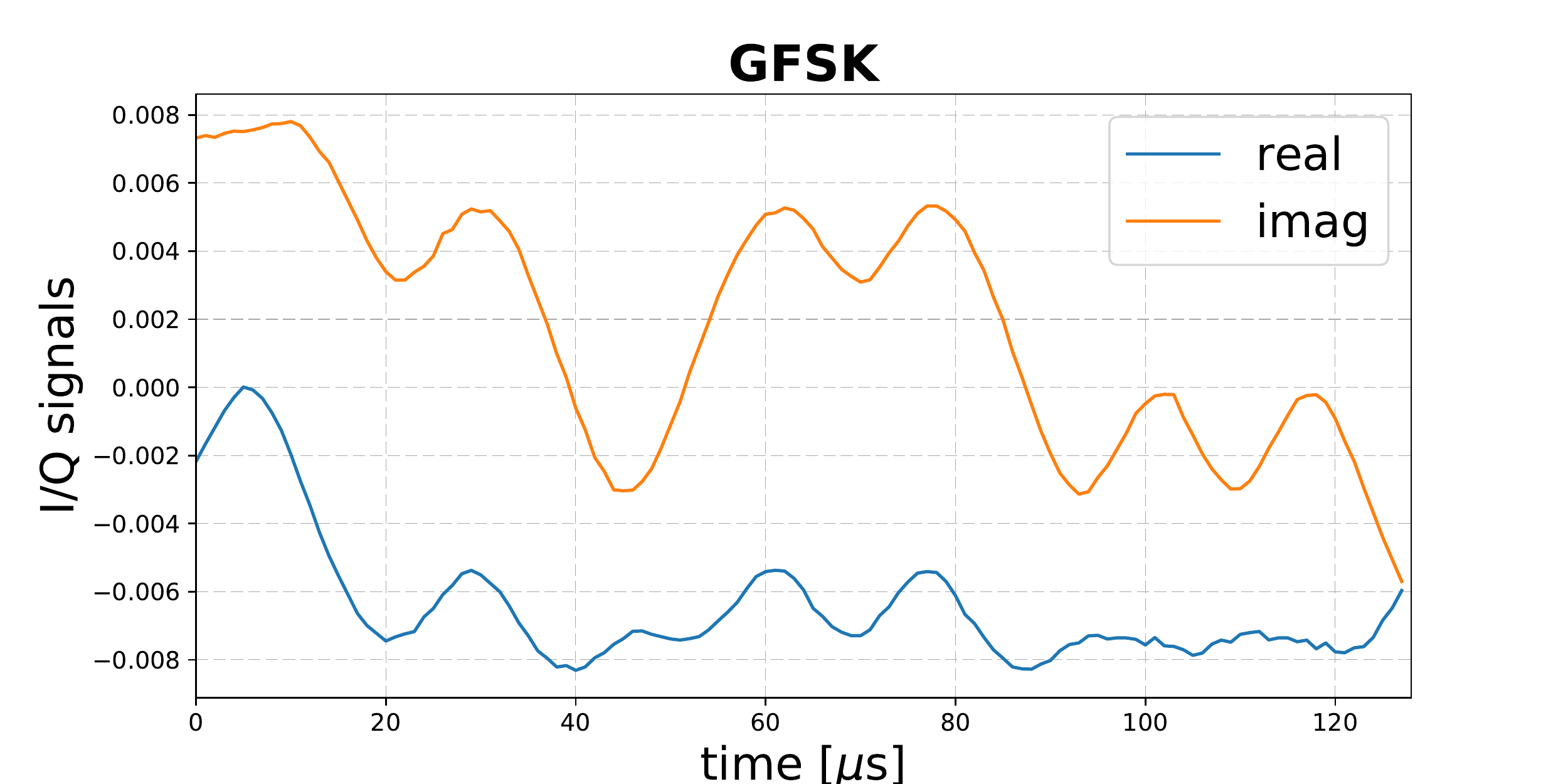}}
\subfloat[\textbf{PAM4}]{\includegraphics[trim={0 0 0 1.4cm},clip,width=0.19\textwidth]{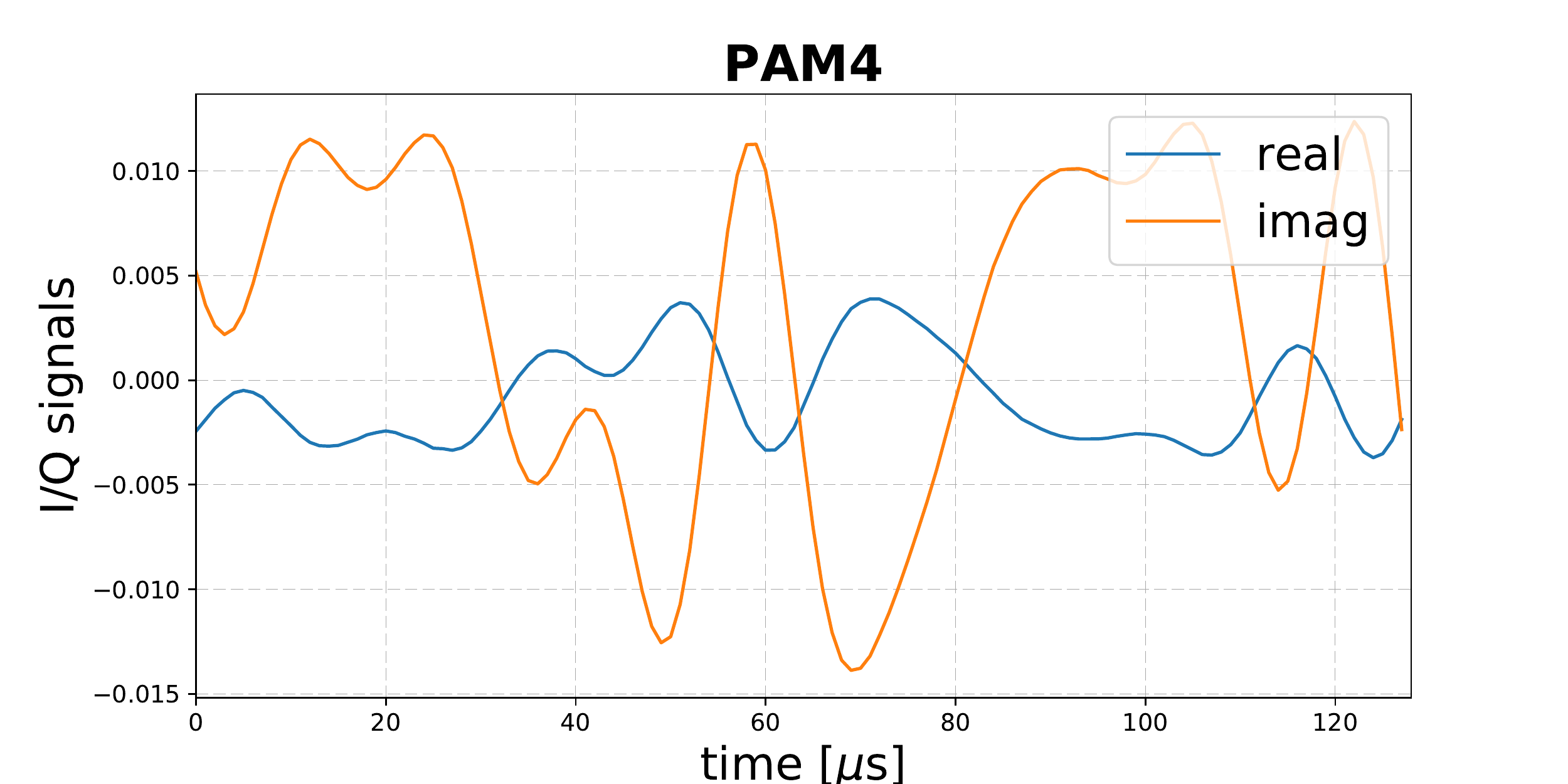}}
\caption{I and Q signals time plot for various modulation schemes} 
\label{fig:IQvisual}        
\end{figure}

\textbf{Transformation 3 (FFT vector)}: \textit{The \textbf{FFT vector} is a mapping from the raw time-domain  complex data vector $\textbf{r}_k \in \mathcal{C}^{N}$ into its frequency-domain representation vector consisting of two sets of real-valued data vectors, one that carries the real component of its complex FFT $\textbf{x}_{F_{re}}$ and one that holds the imaginary component of its FFT $\textbf{x}_{F_{im}}$.} That is
\begin{equation}
\textbf{x}^{\mathcal{F}}_k=
\begin{bmatrix}
	{\textbf{x}_{F_{re}}}^T \\
	{\textbf{x}_{F_{im}}}^T 
\end{bmatrix}
\end{equation}

The translation to frequency-domain is performed by a Fast Fourier Transform (FFT) denoted by $\mathcal{F}$ so that
\begin{align*}
\mathcal{F}: \textbf{r}_k  \mapsto \textbf{w} \\
\textbf{x}_{F_{re}}=\Re\{\textbf{w}\} \\
\textbf{x}_{F_{im}}=\Im\{\textbf{w}\}
\end{align*}

Here, $\textbf{w} \in \mathcal{C}^{N}$, $\textbf{x}_{F_{re}}, \textbf{x}_{F_{im}} \in \mathcal{R}^{N}$ while $\Re\{.\}$ and $\Im\{.\}$ can be conceived as operators giving the real and imaginary parts of a complex vector, respectively. Thus, the resulting \textbf{\textit{FFT vector}} is $\textbf{x}^{\mathcal{F}}_k \in \mathcal{R}^{2xN}$.
In short, this may be denoted as
\begin{align*} 
f: \mathbb{C}^N &\to \mathbb{R}^{2xN} \\
 \textbf{r}_k  &\mapsto \textbf{x}^{\mathcal{F}}_k
\end{align*}

\begin{figure}[t]
\centering
\subfloat[\textbf{BPSK}]{\includegraphics[trim={0 0.5cm 0 1.1cm},clip, width=0.48\textwidth]{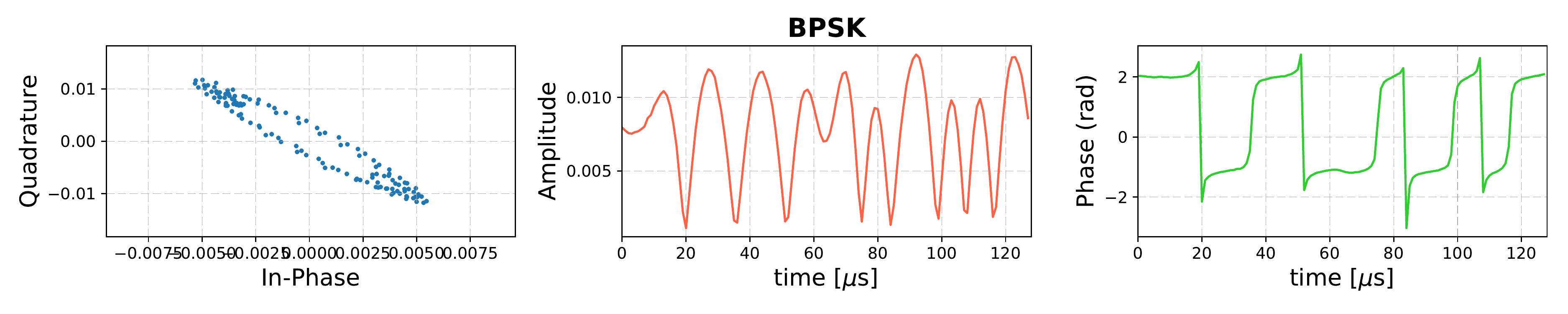}}
\hfil
\subfloat[\textbf{QPSK}]{\includegraphics[trim={0 0.5cm 0 1.1cm},clip, width=0.48\textwidth]{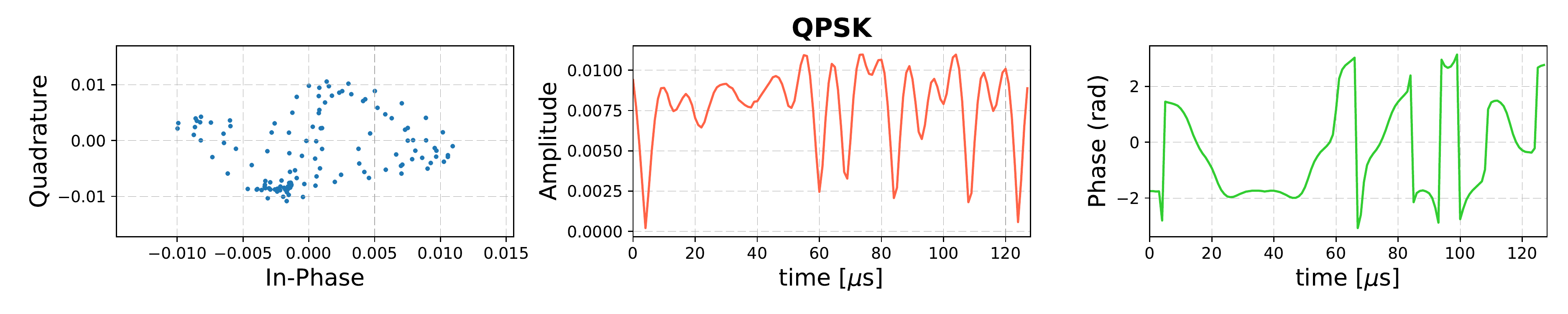}}
\hfil
\subfloat[\textbf{8PSK}]{\includegraphics[trim={0 0.5cm 0 1.1cm},clip, width=0.48\textwidth]{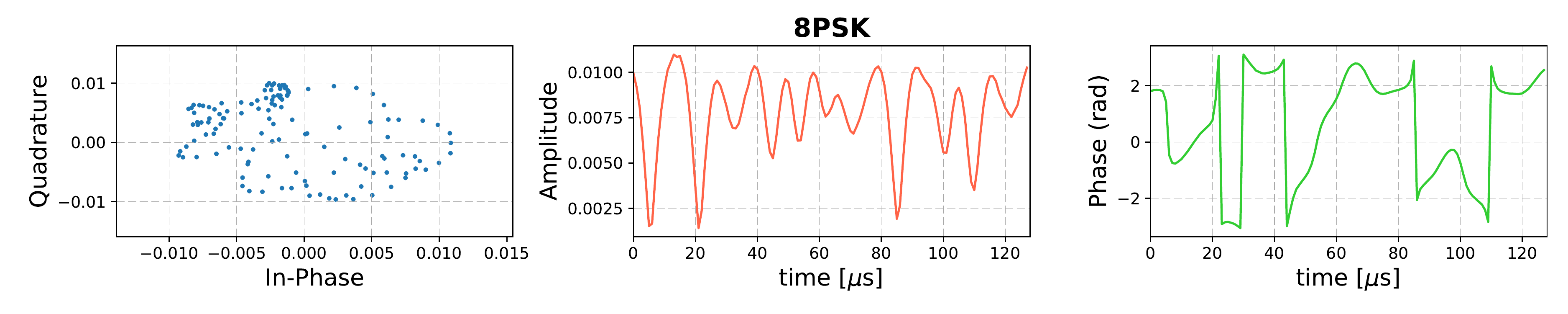}}
\hfil
\subfloat[\textbf{QAM16}]{\includegraphics[trim={0 0.5cm 0 1.1cm},clip, width=0.48\textwidth]{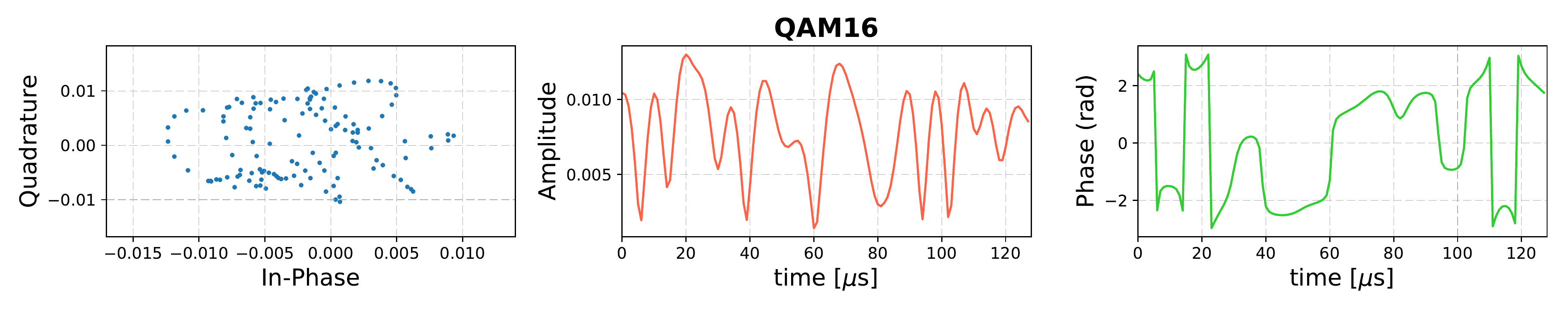}}
\hfill
\subfloat[\textbf{QAM64}]{\includegraphics[trim={0 0.5cm 0 1.1cm},clip, width=0.48\textwidth]{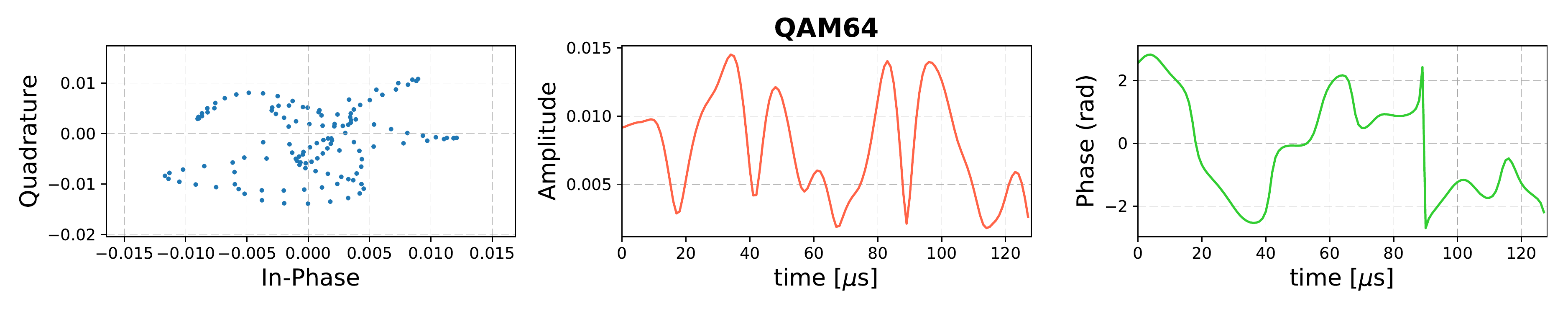}}
\hfill
\subfloat[\textbf{CPFSK}]{\includegraphics[trim={0 0.5cm 0 1.1cm},clip, width=0.48\textwidth]{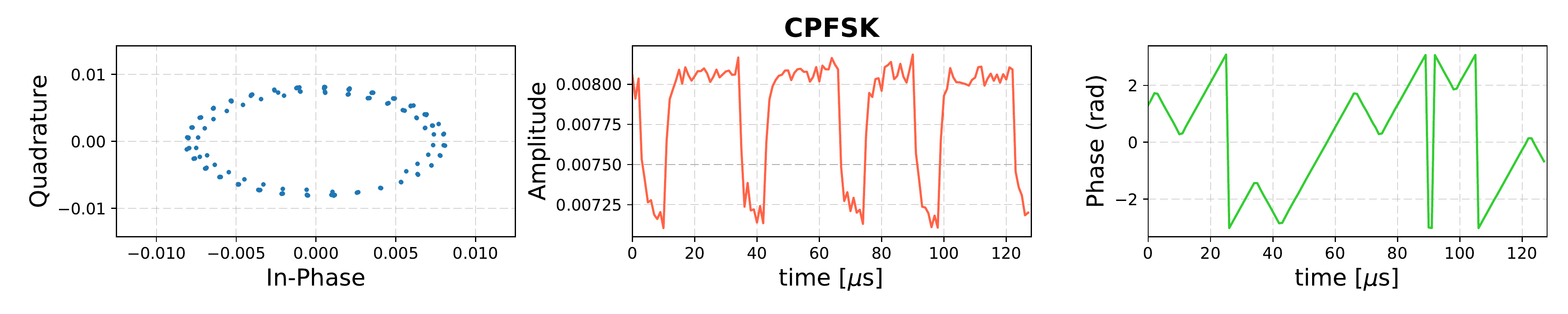}}
\hfill
\subfloat[\textbf{GFSK}]{\includegraphics[trim={0 0.5cm 0 1.1cm},clip, width=0.48\textwidth]{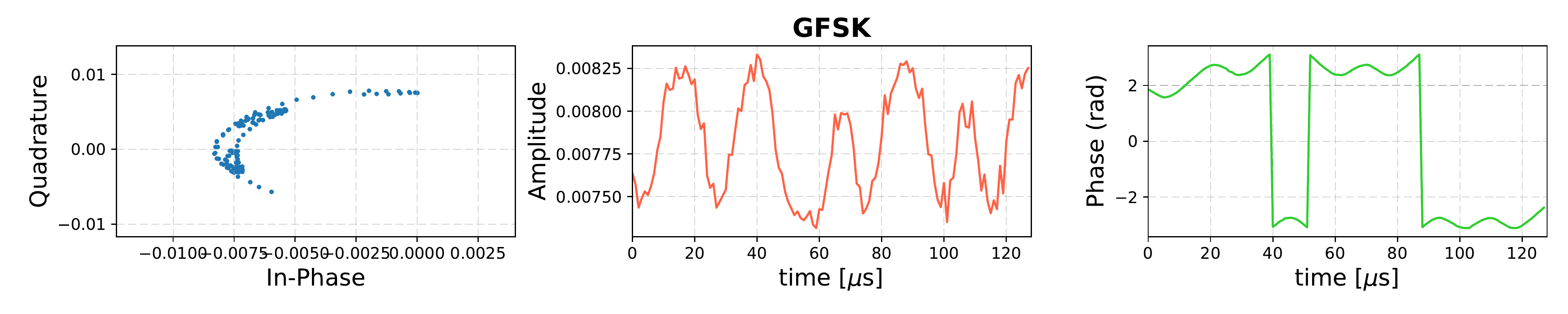}}
\hfill
\subfloat[\textbf{PAM4}]{\includegraphics[trim={0 0.5cm 0 1.1cm},clip, width=0.48\textwidth]{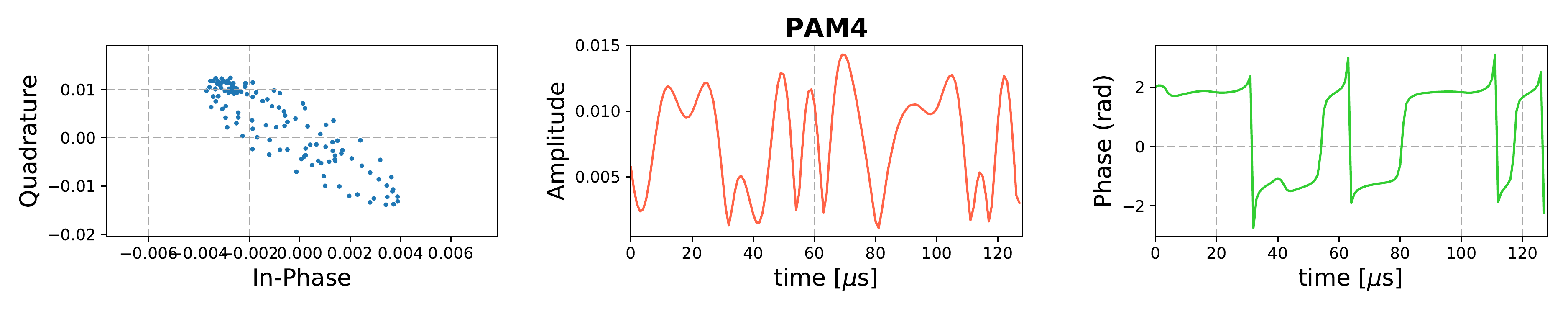}}
\centering
\caption{Constellation diagram, Amplitude and Phase signal time plot for various modulation schemes}  
\label{fig:5}        
\end{figure}

\begin{figure}[tb]
\centering
\subfloat[\textbf{BPSK}]{\includegraphics[trim={0 0 0 1.4cm},clip,width=0.19\textwidth]{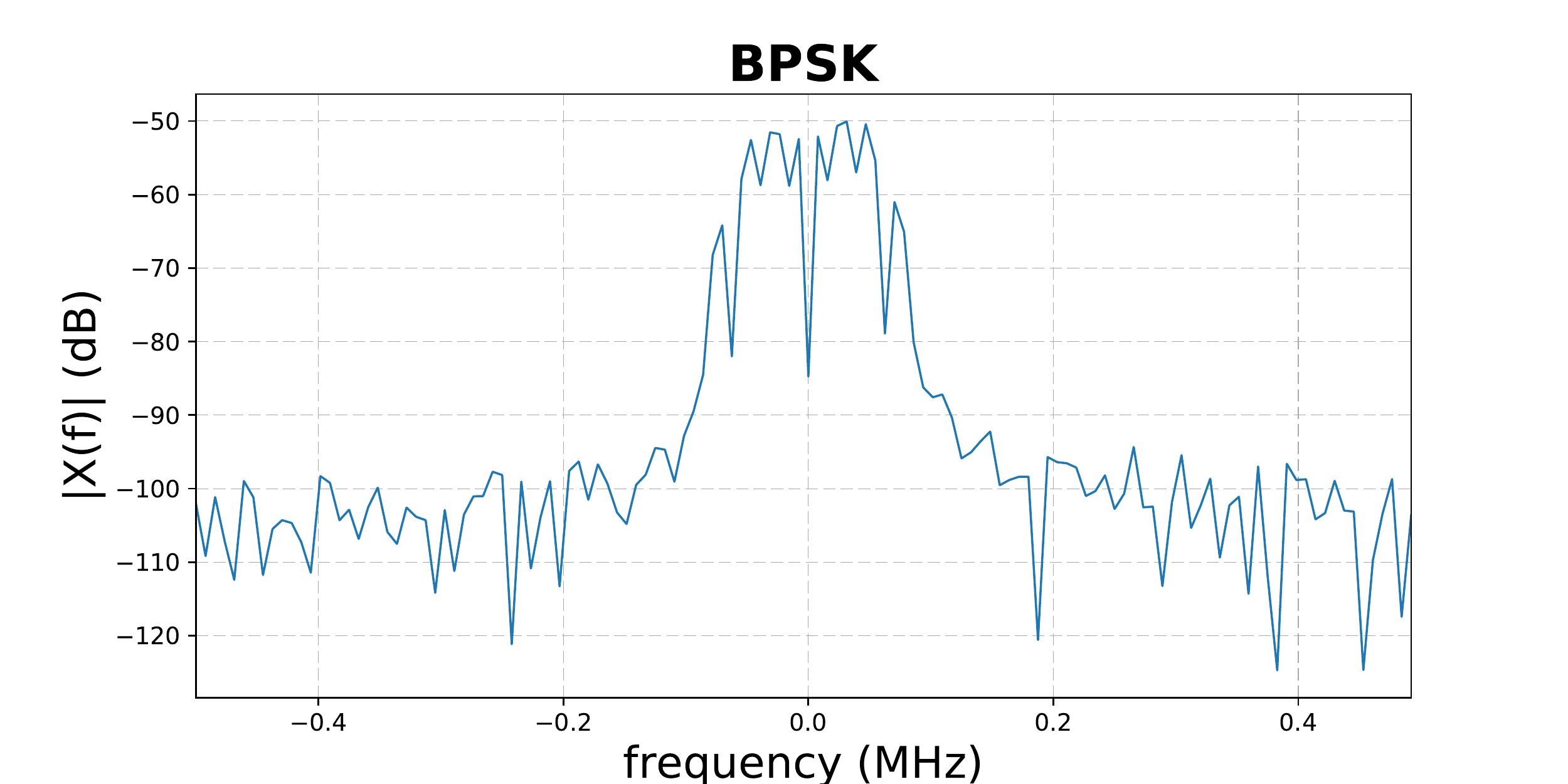}}
\subfloat[\textbf{QPSK}]{\includegraphics[trim={0 0 0 1.4cm},clip,width=0.19\textwidth]{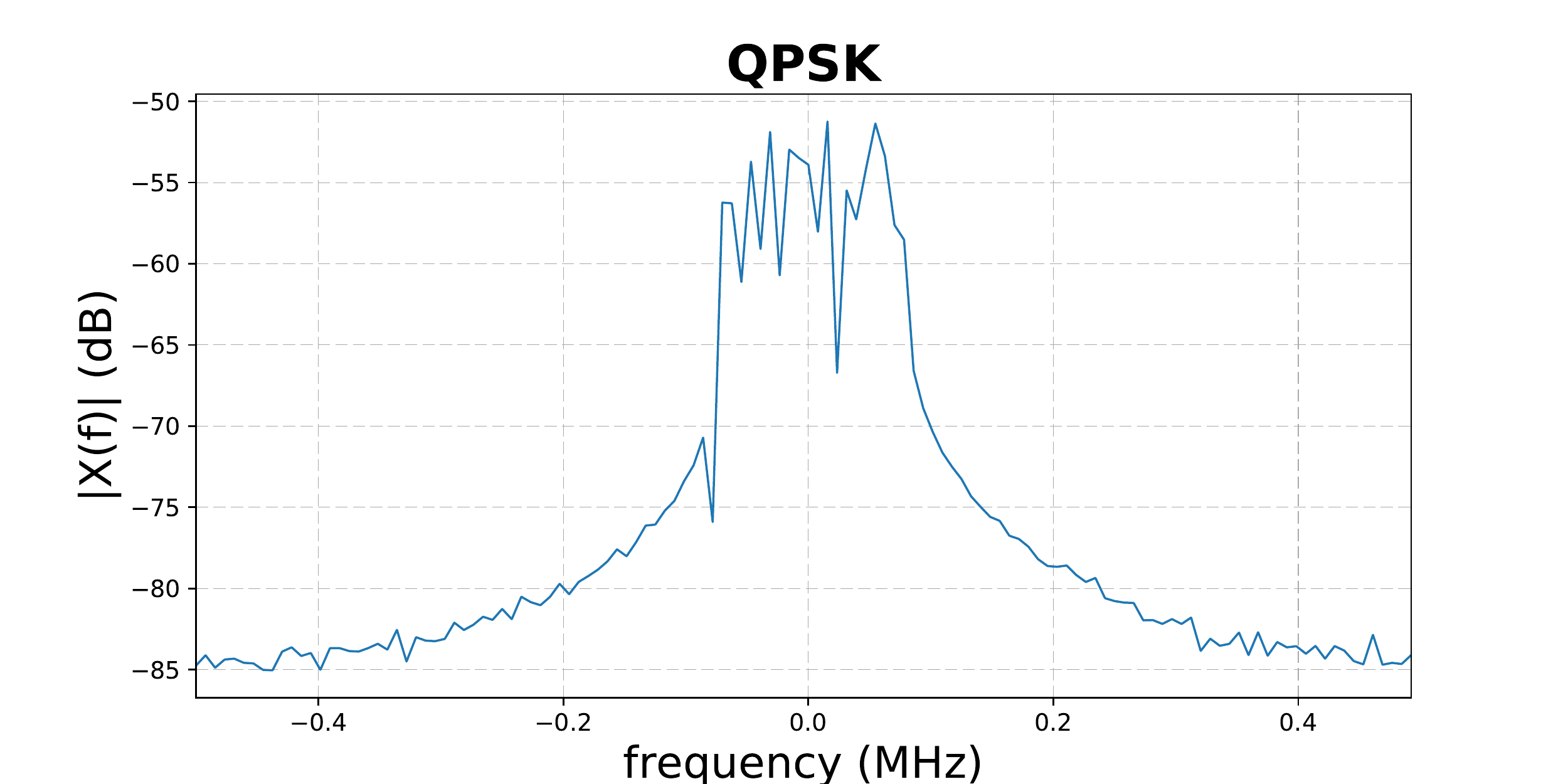}}
\hfil
\subfloat[\textbf{8PSK}]{\includegraphics[trim={0 0 0 1.4cm},clip,width=0.19\textwidth]{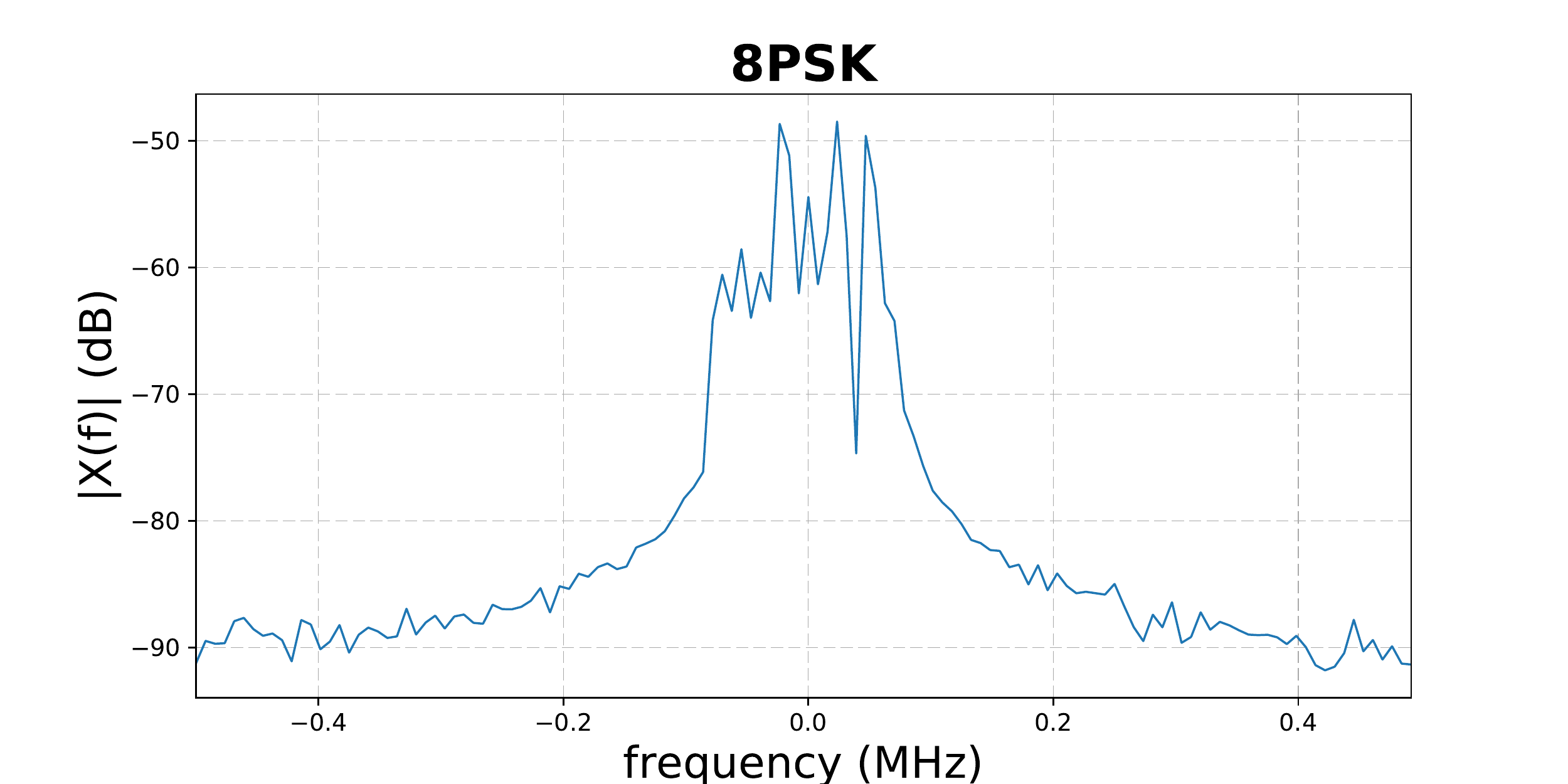}}
\subfloat[\textbf{QAM16}]{\includegraphics[trim={0 0 0 1.4cm},clip,width=0.19\textwidth]{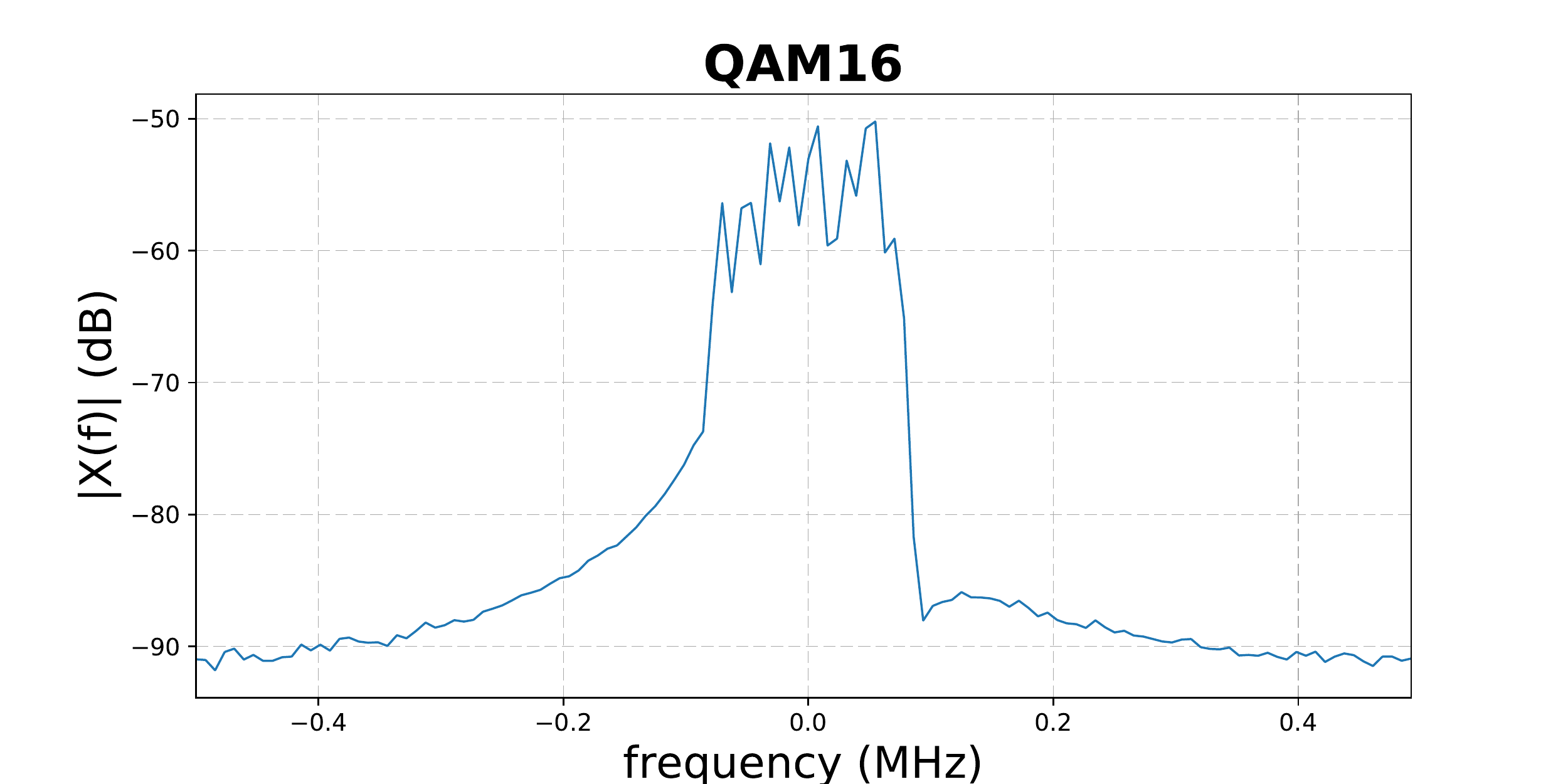}}
\hfill
\subfloat[\textbf{QAM64}]{\includegraphics[trim={0 0 0 1.4cm},clip,width=0.19\textwidth]{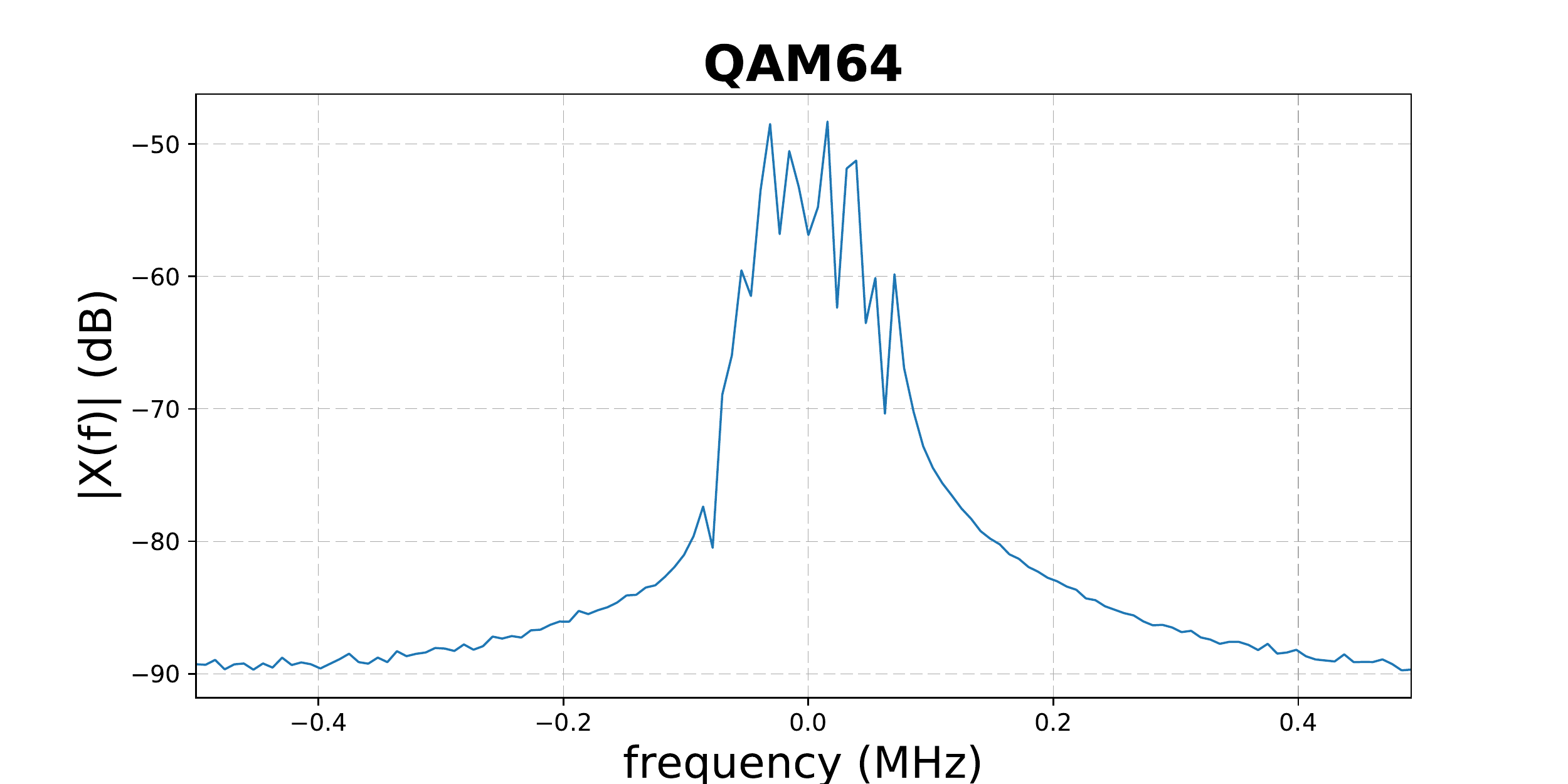}}
\subfloat[\textbf{CPFSK}]{\includegraphics[trim={0 0 0 1.4cm},clip,width=0.19\textwidth]{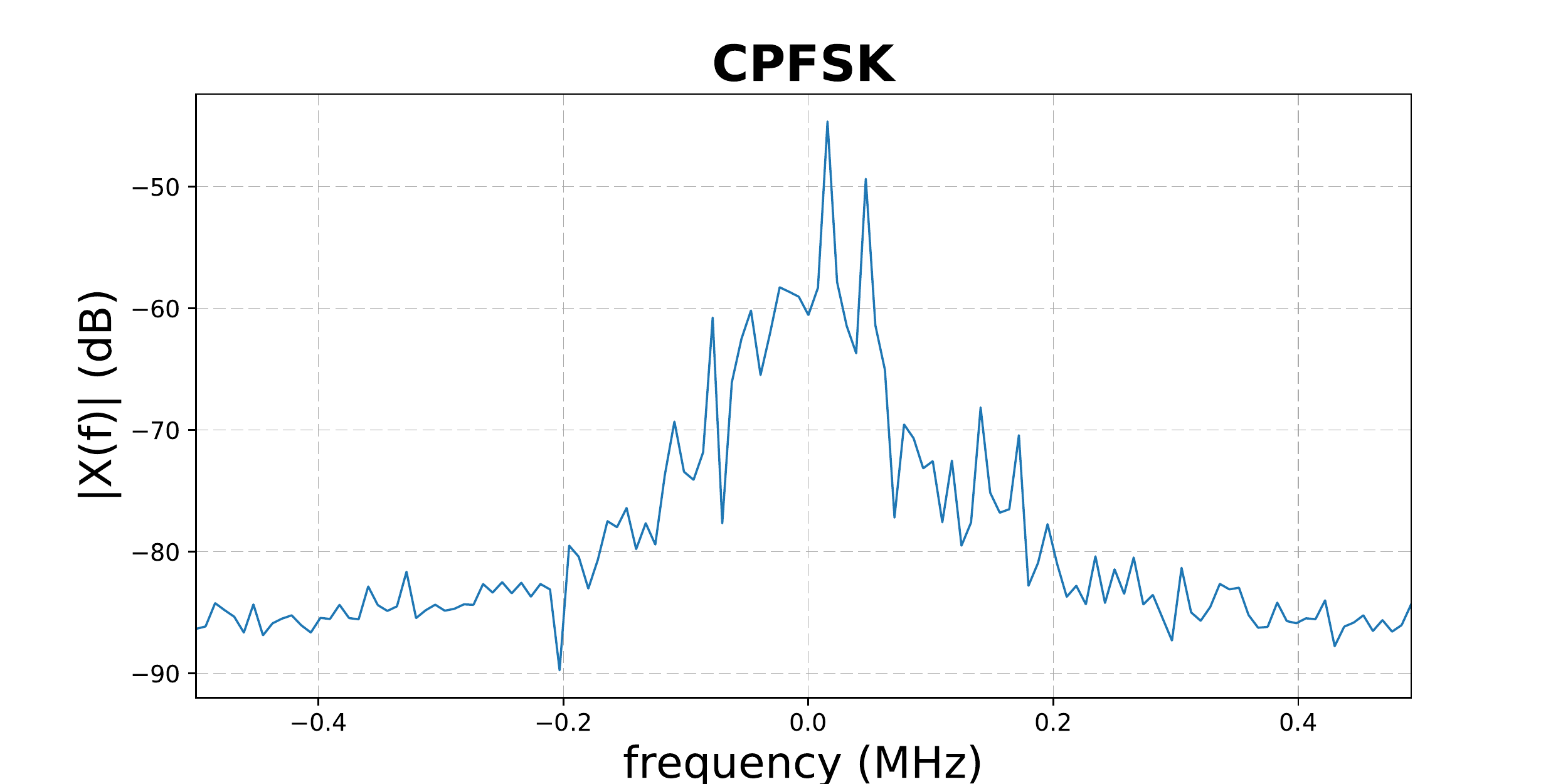}}
\hfill
\subfloat[\textbf{GFSK}]{\includegraphics[trim={0 0 0 1.4cm},clip,width=0.19\textwidth]{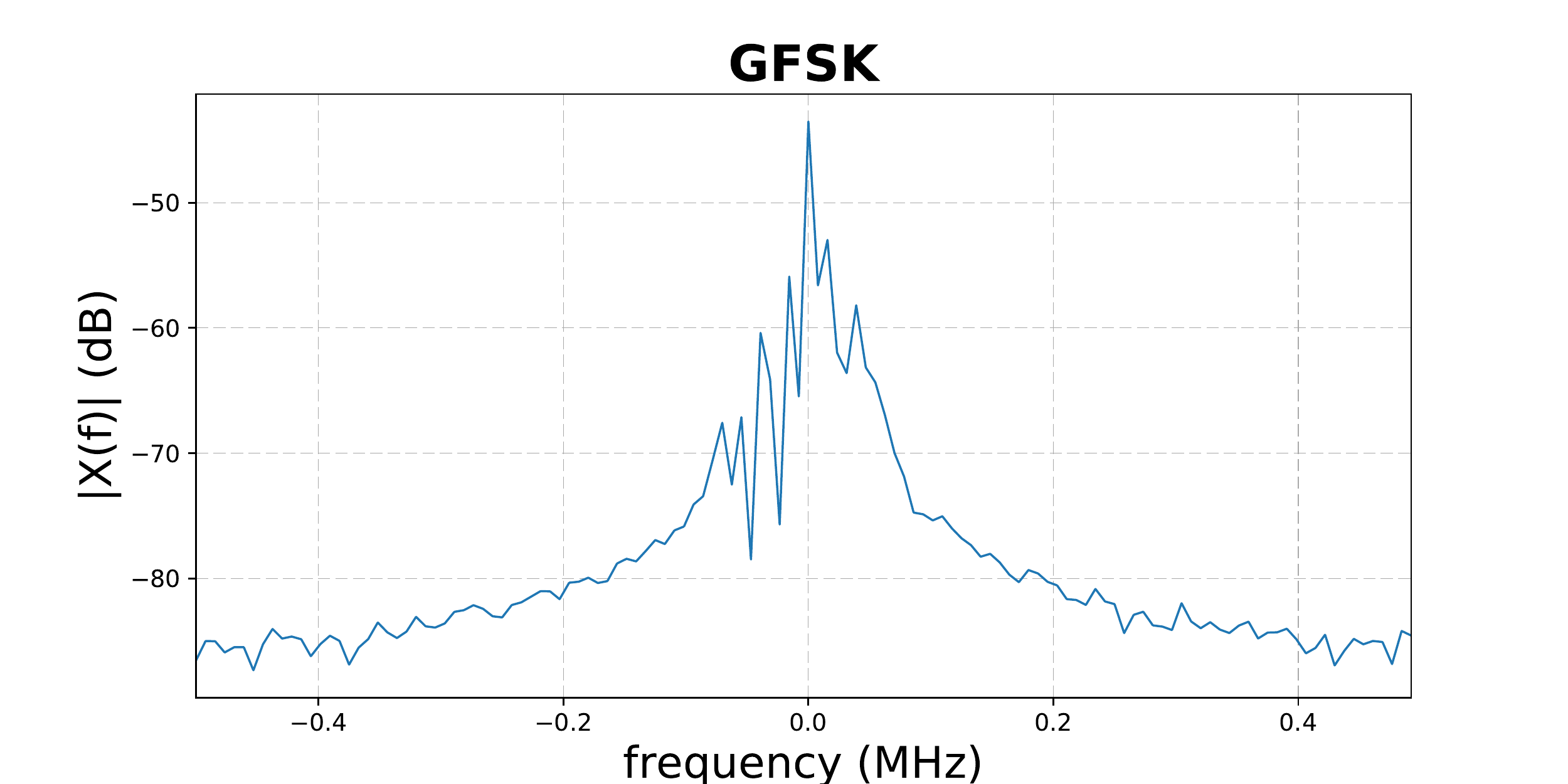}}
\subfloat[\textbf{PAM4}]{\includegraphics[trim={0 0 0 1.4cm},clip,width=0.19\textwidth]{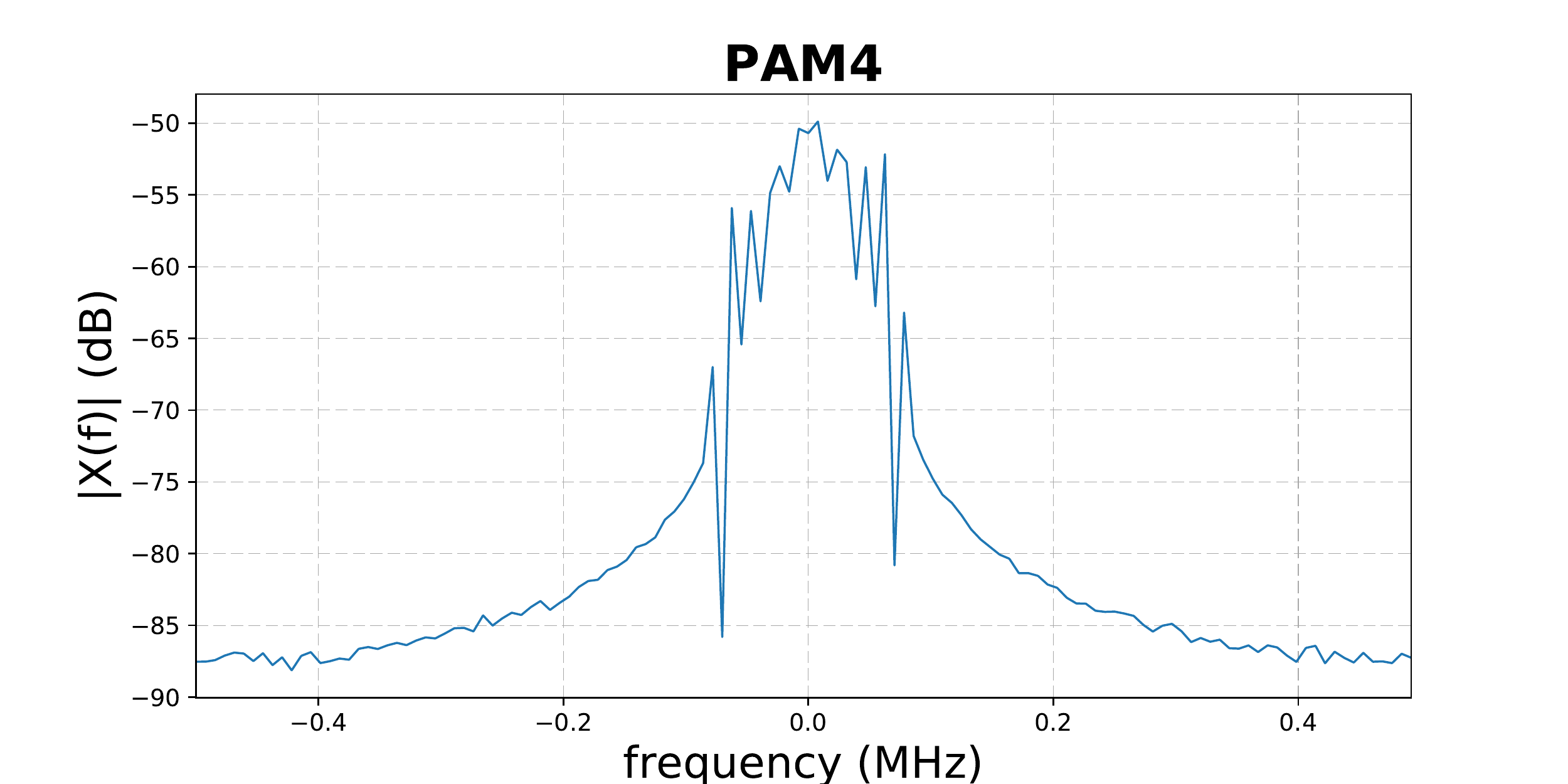}}
\caption{Frequency magnitude spectrum for various modulation schemes} 
\label{fig:fftvisual}        
\end{figure}

Figures \ref{fig:IQvisual}, \ref{fig:5} and \ref{fig:fftvisual} visualize examples of $\textbf{IQ}$, \textbf{$\textbf{A}/\boldsymbol{\upphi}$} and \textbf{FFT} feature vectors, respectively. The visualizations show representations for different modulation formats passed through a channel model with impairments as described in \ref{radiosig}. These are examples of 128 samples
for modulation formats depicted from the "RadioML Modulation" dataset introduced in Section \ref{sec:datasets}.
Figure \ref{fig:IQvisual} shows $\textbf{x}^{IQ}_k$ time plots of the raw sampled complex signal at the receiver for different modulation types.
Figure \ref{fig:5} shows the amplitude and phase time plots for modulation format examples.
Figure \ref{fig:fftvisual} shows their frequency magnitude spectrum. It can be seen that the signals are corrupted due to the wireless channel effects and transmitter-receiver synchronization imperfections, but there are still distinctive patterns that can be used for deep learning to extract high level features for wireless signal identification.

The motivation behind using these three transformations is to train three deep learning models where: one will explore the raw data to discover the patterns and temporal features solely from raw samples, one will see the amplitude and phase information in the time domain, while the third will see the frequency domain representation  to perform feature extraction in the frequency space.
We investigate how the choice of data representation influences the classification accuracy.
The data representations have been carefully designed so that all of them create a vector of the same dimension and type in $\mathcal{R}^{2xN} $. The reason for that is to obtain a unified vector shape which will allow to use the same CNN architecture for training on all three data representations and for different use cases.

\subsection{Wireless signal classification}

The problem of identifying the wireless signals from spectrum data can be treated as a data-driven machine learning \textit{classification} problem. In order to apply ML techniques to this setup, as described in Section \ref{ml} the wireless communication problem has to be formulated as a parametric estimation problem where certain parameters are unknown and need to be estimated.

Given a set of \textit{K} wireless signals to be detected, the problem of identifying a signal from this set turns into a \textit{K}-class classification problem.
Suppose a data measurement point knows the transmitted signal type (e.g. modulation type, interfering emitter type, etc.)
for a time period $t=[0,T)$ (i.e. a "training period")
and collects several complex baseband time series of $n$ measurements for each signal type into a data vector $\textbf{r}_k$, as described in Section \ref{sec:dataAq}. In total, $m$ snapshots for the data vectors $\textbf{r}_k$ are collected.
These data vectors contain emitting signals that contain distinctive features. In order to extract these features, each data vector is transformed into a feature vector, $\textbf{x}_k$, according to the data transformations introduced in Section \ref{sec:sigTransf} and the results are stacked into an observation matrix $\textbf{X} \in \mathcal{R}^{mxn}$. Each data vector is further annotated with the corresponding wireless signal type in form of a discrete one-hot encoded vector $\textbf{y}_k \in \mathcal{R}^K$, $k=1,...,m$.

The obtained data pairs, $\{(\textbf{x}_k, \textbf{y}_k), k=1,...,m \}$, form a dataset suitable to estimate the parameters, $\boldsymbol{\uptheta}$, that characterize the wireless signal classifier, $f$. 

It is instructive to note that the training phase presumes \textit{a prior} information about the type of wireless signal the was used on the transmitter. However, once the classifier is trained this information will no longer be necessary and the signals may be automatically identified by the model. That is, for the $i$-th spectrum data vector input, $\textbf{x}_i$, the predictor's last layer can automatically output an estimate of the probability $P(y_i = k|x_i; \theta)$, where $k$ ranges from $0$ to $K-1$. That is a score class. Finally, the predicted class is then the one with the highest score, i.e. $\hat{y_i}=\operatornamewithlimits{argmax}\limits_{k}{P(y_i = k|x_i; \theta)}$.

\section{Evaluation Setup}
\label{sec:evaluation}
To evaluate end-to-end learning from spectrum data, we train  CNN wireless signal classifiers for two use cases: (i) Radio signal modulation recognition and (ii) Wireless interference identification, for different wireless data representations.

\textbf{Radio signal modulation recognition} relates to the problem of identifying the modulation structure of the received wireless signal in spectrum monitoring tasks, as a step towards understanding what type of communication scheme and emitter is present. Modulation recognition is vital for radio spectrum regulation and in dynamic spectrum access applications.

\textbf{Wireless interference identification} is the task of identifying the type of coexisting wireless emitter,  that is operating in the same frequency band. This is essential for effective interference mitigation and coexistence management in unlicensed frequency bands such as, for example, the 2.4GHz ISM band shared by heterogeneous wireless communication systems.

For each task the CNNs were trained on three characteristic data representations: IQ vectors, Amplitude/Phase vectors and FFT vectors, as introduced in Section \ref{sec:sigTransf}. 
As a result for each task three datasets, $S$, one per data transformation are created. That is,
\begin{align} \label{eq:dataset1}
S^{IQ}=\{(\textbf{x}^{IQ}_k, \textbf{y}_k), k=1,...,m \} \\
S^{\textbf{A}/\boldsymbol{\upphi}}=\{(\textbf{x}^{\textbf{A}/\boldsymbol{\upphi}}_k, \textbf{y}_k), k=1,...,m \} \\
S^{\mathcal{F}}=\{(\textbf{x}^{\mathcal{F}}_k, \textbf{y}_k), k=1,...,m \} \label{eq:dataset3}
\end{align}
where $m$ has the order of tens of thousands instances.

\subsection{Datasets description}
\label{sec:datasets}
\subsubsection{\textbf{Radio Modulation recognition}}
To evaluate end-to-end learning for radio modulation type identification, we consider measurements of the received wireless signal for various modulation formats from the "RadioML 2016.10a Modulation" dataset \cite{o2016convolutional}. Specifically, for all experiments performed in this paper we used labelled data vectors for the following digital modulation formats: BPSK, QPSK, 8-PSK, 16-QAM, 64-QAM, CPFSK, GFSK, 4-PAM, WBFM, AM-DSB, AM-SSB. The data vectors, $\textbf{x}_k$, were collected at a sampling rate $1MS/s$ in $N=128$ sample batches, each containing between 8 and 16 symbols corrupted by random noise, time offset, phase, and wireless channel distortions as described by the channel model in \ref{radiosig}.
One-hot encoding is used to create a discrete set of 11 class labels corresponding to 11 considered modulations, so that the response variable forms a binary 11-vector $\textbf{y}_k \in \mathcal{R}^{11}$. 
The task of modulation recognition is then a 11-class classification problem.
In total, 220,000 data vectors $\textbf{x}_k \in \mathcal{R}^{2x128}$ consisting of I and Q samples are used.

\subsubsection{\textbf{Wireless Interference identification in ISM bands}}

The rise of heterogeneous wireless technologies operating in the unlicensed ISM bands has caused severe communication challenges due to cross-technology interference, which adversely affects the performance of wireless networks.
To tackle these challenges novel agile methods that can assess the channel conditions are needed.
We showcase end-to-end learning as a promising approach that can determine whether communication is feasible over the wireless link by accurately identifying cross-technology interference. Specifically, the "Wireless interference" dataset \cite{schmidt2017wireless} is used which consists of measurements gathered from standardized wireless communication systems based on IEEE 802.11b/g (WiFi), IEEE 802.15.4 (Zigbee) and IEEE 802.15.1 (Bluetooth) standards, operating in the 2.4GHz frequency band.
The dataset is labelled according to the allocated frequency channel and the corresponding wireless technology, resulting in 15 different classes.
Compared to the modulation recognition dataset, this dataset consists of measurements gathered assuming a communication channel model with less channel impairments. In particular, a flat fading channel with additive white Gaussian noise was assumed. I and Q samples were collected 
at a sampling rate $10MS/s$ in batches of 128 each, capturing hereby 1 to 12 symbols for each utilized wireless technology depending on the symbol duration.
In total, 225,225 snapshots were collected.

\subsection{CNN network structure}
The convolutional neural network structure utilized for end-to-end learning from spectrum data is derived from O'Shea at al. \cite{o2016convolutional}, i.e the CNN2 network, as it has shown to significantly outperform traditional signal identification approaches.

\setlength{\tabcolsep}{0.5em} 
{\renewcommand{\arraystretch}{1.2}
\begin{table}
\caption{CNN structure}
\resizebox{\columnwidth}{!}{%
\centering
\begin{tabular*}{0.5\textwidth}{m{1.8cm} c m{1.6cm} c}
\toprule
 \textbf{Layer type} & \textbf{Input size} & \textbf{Parameters} & \textbf{Activation function}\\    \hline
Convolutional layer & 2x128 & 256 filters, filter size 1x3, dropout=0.6 & ReLU\\   \hline
Convolutional layer & 256x2x128 & 80 filters, filter size 2x3, dropout=0.6 & ReLU\\ \hline  
Fully connected layer & 10240x1 & 256 neurons, dropout=0.6 & ReLU\\    \hline
Fully connected layer & 256x1 & 11 neurons or 15 neurons & Softmax\\  
\bottomrule
\end{tabular*}
}
\label{cnnstructure}
\end{table}

Table \ref{cnnstructure} provides a summary of the utilized CNN network. The visible layer of the network has a unified size of $2x128$ receiving either \textbf{IQ}, \textbf{FFT} or \textbf{Amplitude/Phase} captured data vectors, $\textbf{x}_k \in \mathcal{R}^{2x128}$, that contain sample values of the complex wireless signal. Two hidden convolutional layers further extract high-level features from the input wireless signal representation using kernels and ReLU activation functions. The first convolutional layer consists of 256 stacked filters of size $1x3$ that perform a 2D convolution on the input complex signal representation padded such that the output has the same length as the original input. These filters generate 256 ($2x128$) feature maps that are fed as input to the second layer which has 80 filters of size $2x3$.
To reduce overfitting, in each layer regularization is used with a Dropout $p=0.6$.
Finally, a fully connected layer with 256 neurons and ReLU units is added. The output of this layer is fed to a softmax classifier that estimates the likelihood of the input signal, $x$, belonging to a particular class, $y$. That is $P(y=k|x;\theta)$, where $k$ is a one-hot encoded vector so that $k \in \mathcal{R}^{15}$ for the wireless interference identification case, and $k \in \mathcal{R}^{11}$ for modulation recognition. 

\subsection{Implementation details}
The CNNs were trained and validated using the Keras \cite{chollet2015keras} library on a high computation platform on Amazon Elastic Compute (EC) Cloud with the central processing unit (CPU) Intel(R) Xeon(R) CPU E5-2686 v4 @ 2.30GHz, with 60GB RAM and the Cuda enabled graphics processing unit (GPU) Nvidia Tesla K80. 
For both use cases, 67\% randomly selected examples are used for training in batch sizes of 1024, and 33\% for testing and validation. Hence, for modulation recognition 147,400 examples are used for training, while 72,600 examples for testing and validation. For the task of interference identification, 151,200 examples are training examples, while 74,025 examples are used to test the model.
Both sets of examples are uniformly distributed in Signal-to-Noise Ratio (SNR) from -20dB to +20dB and tagged so that performance can be evaluated on specific subsets. To estimate the model parameters the Adaptive moment estimation (Adam) optimizer \cite{kingma2014adam} was used with a learning rate $\alpha=0.001$, and the input data was normalized to speed up convergence of the learning algorithm.
The CNNs were trained on 70 epochs and the model with the lowest validation loss is selected for evaluation. In total, 6 CNNs were trained, i.e. one for each use case and signal representation. Three for modulation recognition: CNN$^M_{IQ}$, CNN$^M_{A/\upphi}$ and CNN$^M_{\mathcal{F}}$, and three for technology identification CNN$^{IF}_{IQ}$, CNN$^{IF}_{A/\upphi}$ and CNN$^{IF}_{\mathcal{F}}$.
The training time on the GPU resulted in a duration of approximately 60$s$ per epoch for the CNNs performing interference identification, while 42$s$ for the modulation recognition CNNs.

\subsection{Performance metrics}
In order to characterize and compare the prediction accuracy of the end-to-end wireless signal classification models that recognize modulation type or identify interference, we need to measure how well their predictions match the true response value of the observed spectrum data.
Therefore, the performance of the end-to-end signal classification methods can be quantified by means of the prediction accuracy on a test data sample. If the true value and the estimate of the signal classifiers for any instance $i$ are given by $y_i$ and $\hat{y}_i$, respectively, then the overall classification test error over $m_{test}$ testing snapshots can be defined in the following way:
\begin{equation}
E_{test}=\frac{1}{m_{test}}\sum_{i=1}^{m_{test}} l(\hat{y}_i, y_i)
\end{equation}
The \textit{classification accuracy} is then obtained with $1-E_{test}$.

Furthermore, for each signal snapshot in the test set, intermediate statistics, i.e. the number of
\textit{true positive} (TP), \textit{false positive} (FP) and \textit{false negative} (FN) are calculated as follows:

\begin{itemize}
\item If a signal is detected as being from a particular class and it is also annotated as such in the labelled test data, that instance is regarded as \textit{TP}.

\item If a signal is predicted as being from a particular class but does \textit{not} belong to that class according to the labelled test data, that instance is regarded as \textit{FP}.

\item If a signal is \textit{not} detected in a particular instance but it is present in that instance in the labelled test data, that instance is regarded as \textit{FN}.
\end{itemize}

The intermediate statistics are accumulated over all instances in the test set and used to derive three further performance metrics \textit{precision} (P), \textit{recall} (R) and \textit{F}$_1$ score:
\begin{equation}
P=\frac{TP}{TP+FN} \text{ ,    } 
R=\frac{TP}{TP+FP}
\end{equation}

\begin{equation}
F_1 score=2 \times \frac{precision \times recall}{precision + recall}
\end{equation}

Precision, recall and F$_1$ score are per-class performance metrics. In order to obtain one measure that quantifies the overall performance of the classifier, multiple per-class performance measures are combined using a prevalence-weighted macro-average across the class metrics, P$_{avg}$, R$_{avg}$ and F$_{1_{avg}}$.
For a detailed overview of the per-class performance the confusion matrix is used.

\subsection{Numerical results}

\subsubsection{Classification performance}

The CNN network described in Table \ref{cnnstructure} is trained on three data representations for two wireless signal identification problems. Table \ref{PerfTable} provides the averaged performance for the six classifiers. That is, the prevalence-weighted macro-average of precision, recall and F$_1$ score under three SNR scenarios, high (SNR=18dB), medium (SNR=0dB) and low (SNR=-8dB). 

We observe that the models for interference classification show better performance compared to the modulation recognition case.
For high SNR conditions, the CNN$^{IF}$ models achieve a P$_{avg}$, R$_{avg}$ and F$_{1_{avg}}$ between 0.98 and 0.99. For medium SNR the metrics are in the range of 0.94 and 0.99, while under low SNR conditions the performance slightly degrades to 0.81-0.90.
The CNN$^{M}$ models show less robustness to varying SNR conditions, and in general achieve lower classification performance for all scenarios.
In particular, under high SNR conditions depending on the used data representation the achieved P$_{avg}$, R$_{avg}$ and F$_{1_{avg}}$ are in the range of 0.67-0.86. For medium SNR, the performance degrades more then for the CNN$^{IF}$ models, with a P$_{avg}$, R$_{avg}$ and F$_{1_{avg}}$ in the range of 0.59-0.75. Under low SNR, the CNN$^{M}$ models show poor performance with the metrics values in the range of 0.22-0.36.

\setlength{\tabcolsep}{0.5em} 
{\renewcommand{\arraystretch}{1.2}
\begin{table}
\centering
\caption{Performance comparison for the trained CNN signal classifier models for three SNR scenarios}
\resizebox{\columnwidth}{!}{%
\begin{tabular}{c c c c c c}
\toprule
\textbf{Case} & \textbf{Model} & \textbf{SNR} & \textbf{P$_{avg}$} & \textbf{R$_{avg}$} & \textbf{$F_1$ score$_{avg}$} \\    \hline
        
\multirow{9}{*}{Mod. recognition} & \multirow{3}{*}{CNN$^M_{IQ}$} & High & 0.83 & 0.82 & 0.79  \\    
															&	 & Medium & 0.75 & 0.75 & 0.72  \\
															&	 & Low & 0.36 & 0.32 & 0.30  \\ \cline{2-6}
								 & \multirow{3}{*}{CNN$^M_{A/\upphi}$} & High & 0.86 & 0.84 & 0.82  \\    
																&	 & Medium & 0.70 & 0.70 & 0.69  \\
																&	 & Low & 0.33 & 0.29 & 0.26  \\ \cline{2-6}
								& \multirow{3}{*}{CNN$^M_{\mathcal{F}}$} & High & 0.71 & 0.68 & 0.67  \\    
																&	 & Medium & 0.63 & 0.6 & 0.59  \\
																&	 & Low & 0.28 & 0.25 & 0.22  \\ 
																
\hline
\multirow{9}{*}{IF identification} & \multirow{3}{*}{CNN$^{IF}_{IQ}$} & High & 0.98 & 0.98 & 0.98  \\    
																&	 & Medium & 0.95 & 0.94 & 0.94  \\
																&	 & Low & 0.84 & 0.82 & 0.81  \\ \cline{2-6}
								 & \multirow{3}{*}{CNN$^{IF}_{A/\upphi}$} & High & 0.99 & 0.99 & 0.99  \\    
																&	 & Medium & 0.98 & 0.98 & 0.98  \\
																&	 & Low & 0.87 & 0.86 & 0.86  \\ \cline{2-6}
								& \multirow{3}{*}{CNN$^{IF}_{\mathcal{F}}$} & High & 0.99 & 0.99 & 0.99  \\    
																&	 & Medium & 0.99 & 0.99 & 0.99  \\
																&	 & Low & 0.90 & 0.89 & 0.89  \\ 													  				         
\bottomrule
\end{tabular}
}
\label{PerfTable}
\end{table}

This may be explained by the different channel models used for generating the datasets for the two case studies, and the type of signals that need to be discriminated in each problem.
For instance, for the IF case a simple channel model with flat fading was considered, while for modulation recognition the channel model was a time-varying multipath fading channel and other transceiver impairments were also taken into account. Hence, the modulation recognition dataset used a more realistic channel model in the data collection process. However, this impacts the classification performance because it is more challenging to design a robust signal classifier for this case compared to the channel condition considered in the IF classification problem. Furthermore, the signals that are classified for IF detection have different characteristics by design. In particular, they use different medium access schemes, channel bandwidth and modulation techniques, which makes it easier for the classifier to differentiate them. In contrast, the selected modulation recognition signals are more similar to each other, because subsets of modulations are based on similar design principles (e.g. all are single carrier modulations).

\begin{figure}[tbh]
\centering
\subfloat[\textbf{CNN$^M_{\mathcal{IQ}}$}]{\includegraphics[width=0.45\textwidth]
{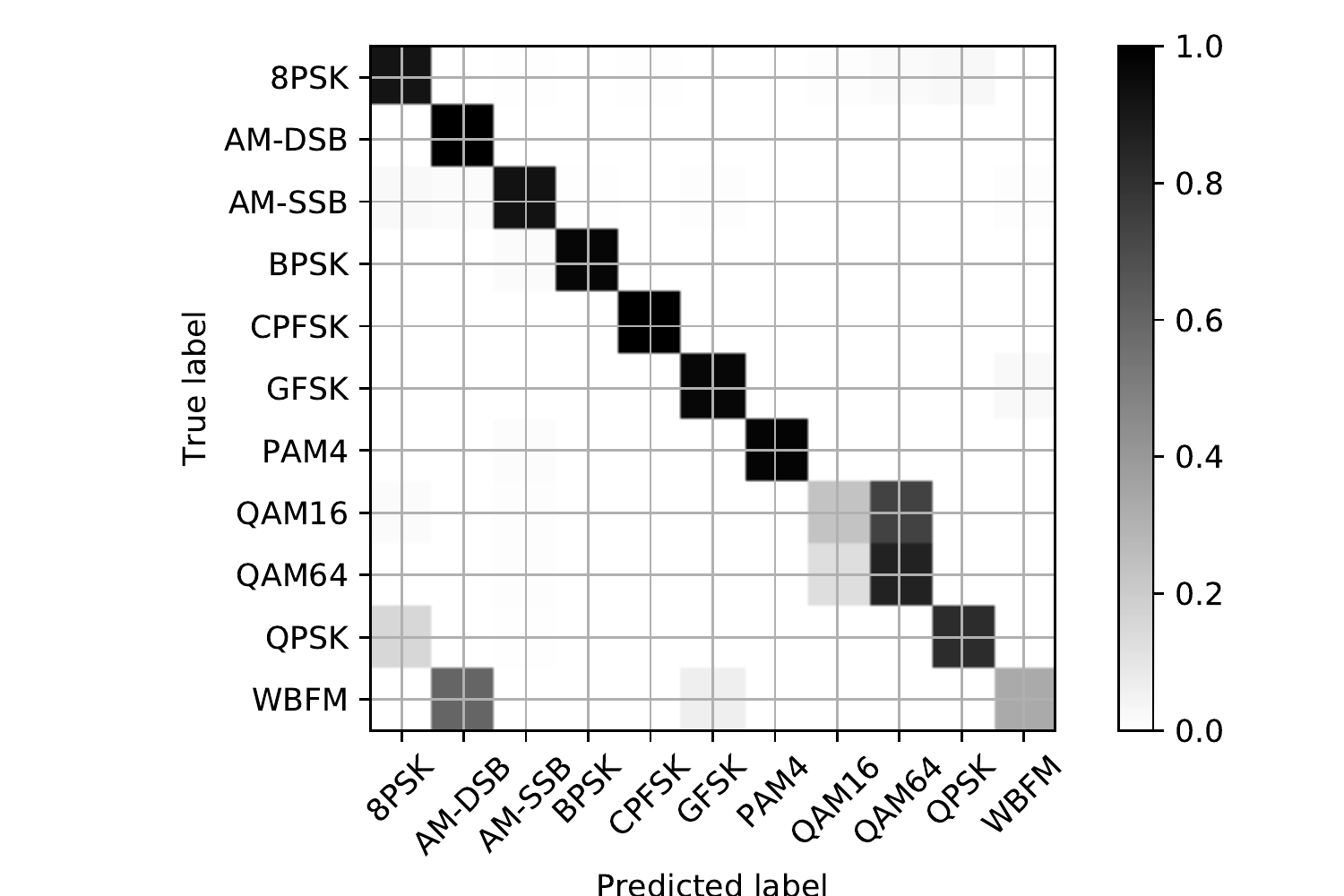}}
\hfil
\subfloat[\textbf{CNN$^M_{\mathcal{\textbf{A}/\boldsymbol{\upphi}}}$}]{\includegraphics[width=0.45\textwidth]
{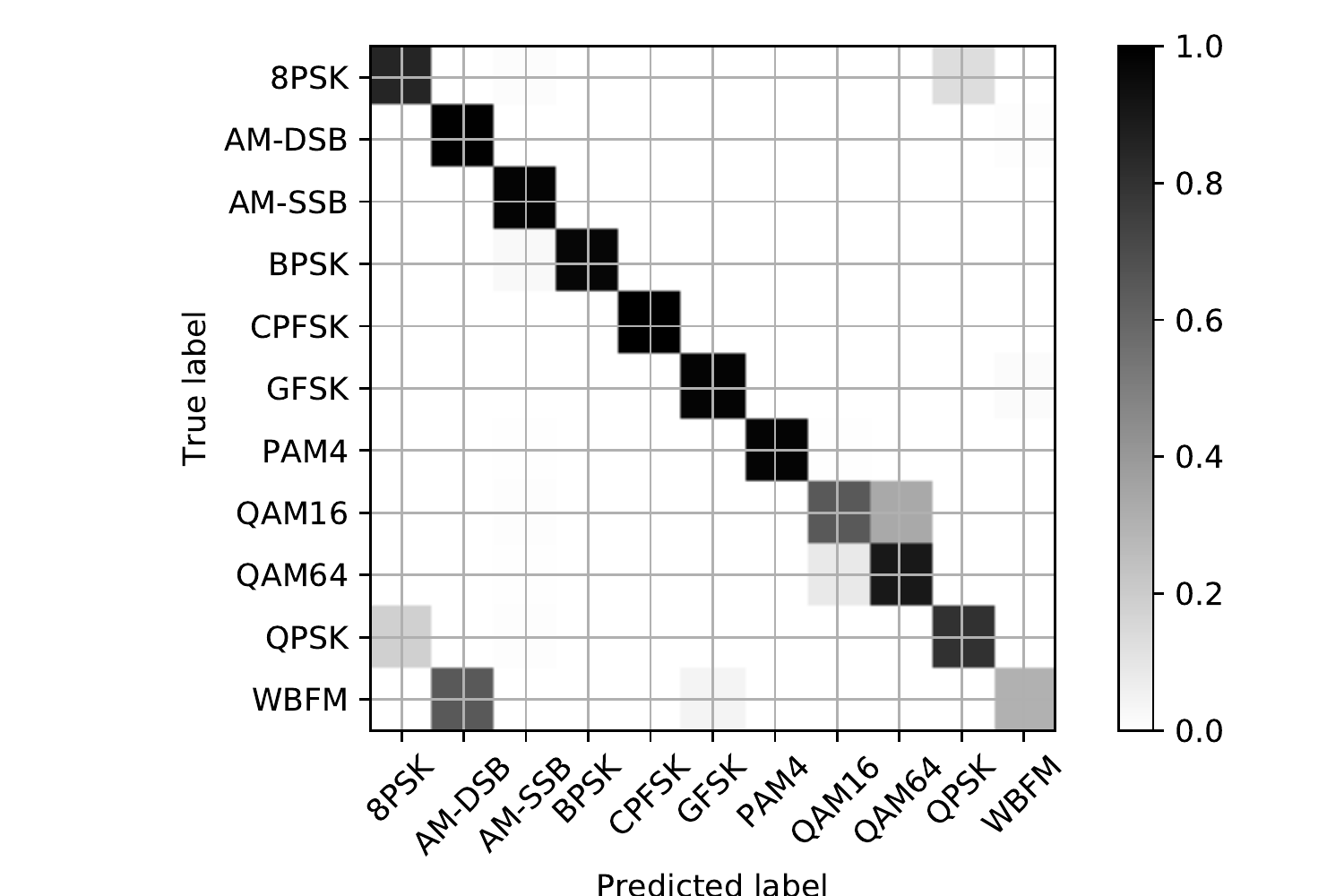}}
\hfil
\subfloat[\textbf{CNN$^M_{\mathcal{F}}$}]{\includegraphics[width=0.45\textwidth]
{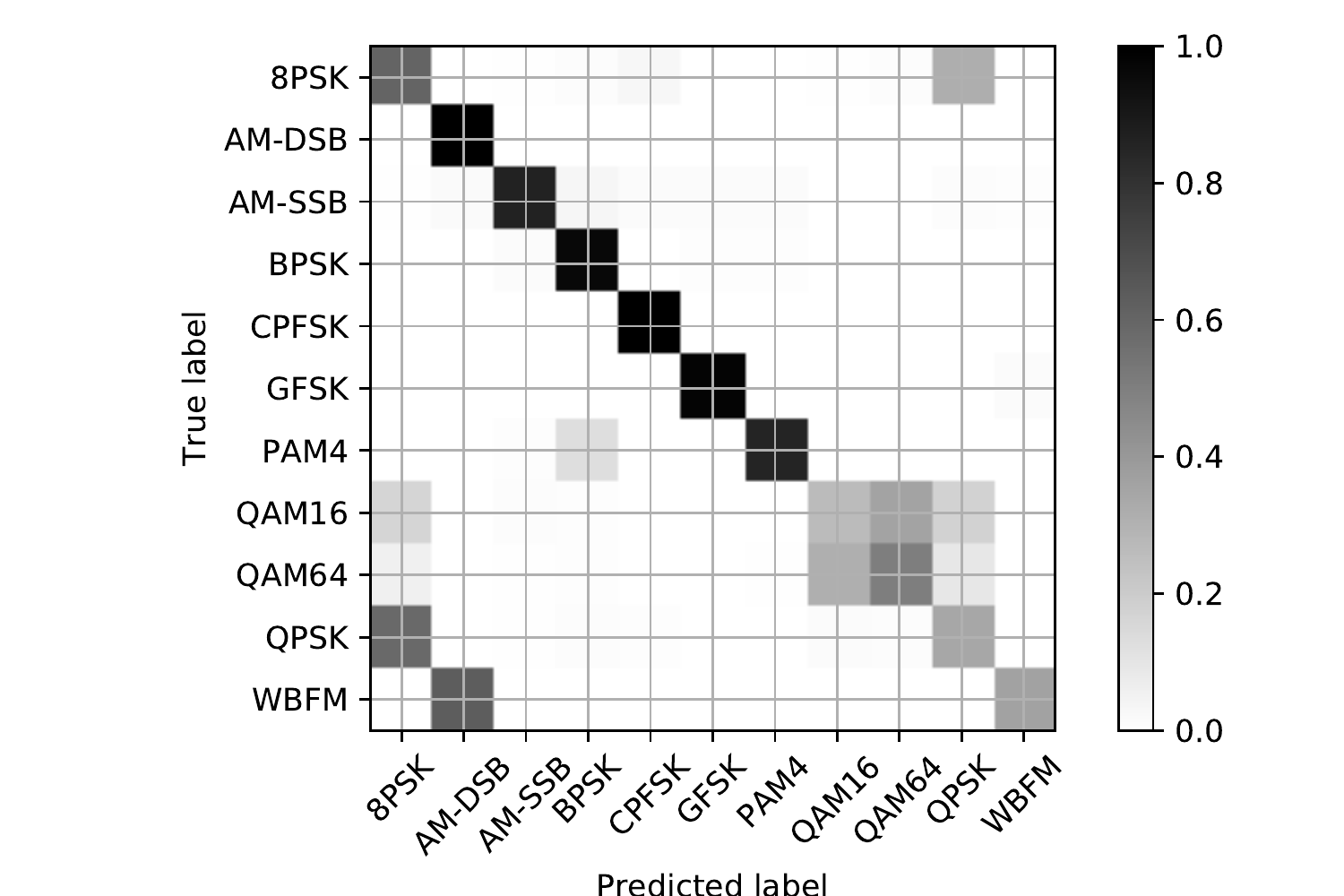}}
\caption{Confusion matrices for the modulation recognition data for SNR 6dB} 
\label{fig:conf_modrec_6dB}   
\end{figure}

To understand the results better confusion matrices for the CNN$^M_{\mathcal{IQ}}$, CNN$^M_{\mathcal{\textbf{A}/\boldsymbol{\upphi}}}$ and 
CNN$^M_{\mathcal{F}}$ models are presented on 
Figure \ref{fig:conf_modrec_6dB} for the case of SNR=6dB.
It can be seen that the classifiers shows good performance by discriminating AM-DSB, AM-SSB, BPSK, CPFSK, GFSK and PAM4 with high accuracy for all three data representations. The main discrepancies are that of QAM16 misclassified as QAM64, which can be explained by the underlying dataset. QAM16 is a subset of QAM64 making it difficult for the classifier to differentiate them. It can be further noticed that the amplitude/phase information helped the model better discriminate QAM16/QAM64, leading to a clearer diagonal for the CNN$^M_{\mathcal{\textbf{A}/\boldsymbol{\upphi}}}$ compared to CNN$^M_{\mathcal{IQ}}$.
There are further difficulties in separating AM-DSB and WBFM signals. 
This confusion may be caused by periods of absence of the signal, as the modulated signals were created from real audio streams.
In case of using the frequency spectrum data, it can be noticed that the CNN$^M_{\mathcal{F}}$ classifier confuses mostly QPSK, 8PSK, QAM16 and QAM16 which is due to their similarities in the frequency domain after channel distortions, making the received symbols indiscernible from each other.


\subsubsection{Noise Sensitivity}
In this section, we evaluate the detection performance for the CNN signal classifiers under different noise levels.
This allows to investigate the communication range over which the classifiers can be effectively used.
To estimate the sensitivity to noise the same testing sets were used labelled with SNR values from -20dB to +20dB and fed into the signal classifiers to obtain the estimated values for each SNR.

Figures \ref{fig:CNN_snr_mod} and \ref{fig:CNN_snr_if} show the obtained results for the modulation recognition and IF identification models, respectively.

\begin{figure}[!t]
    \centering  
    \includegraphics[width=0.5\textwidth]{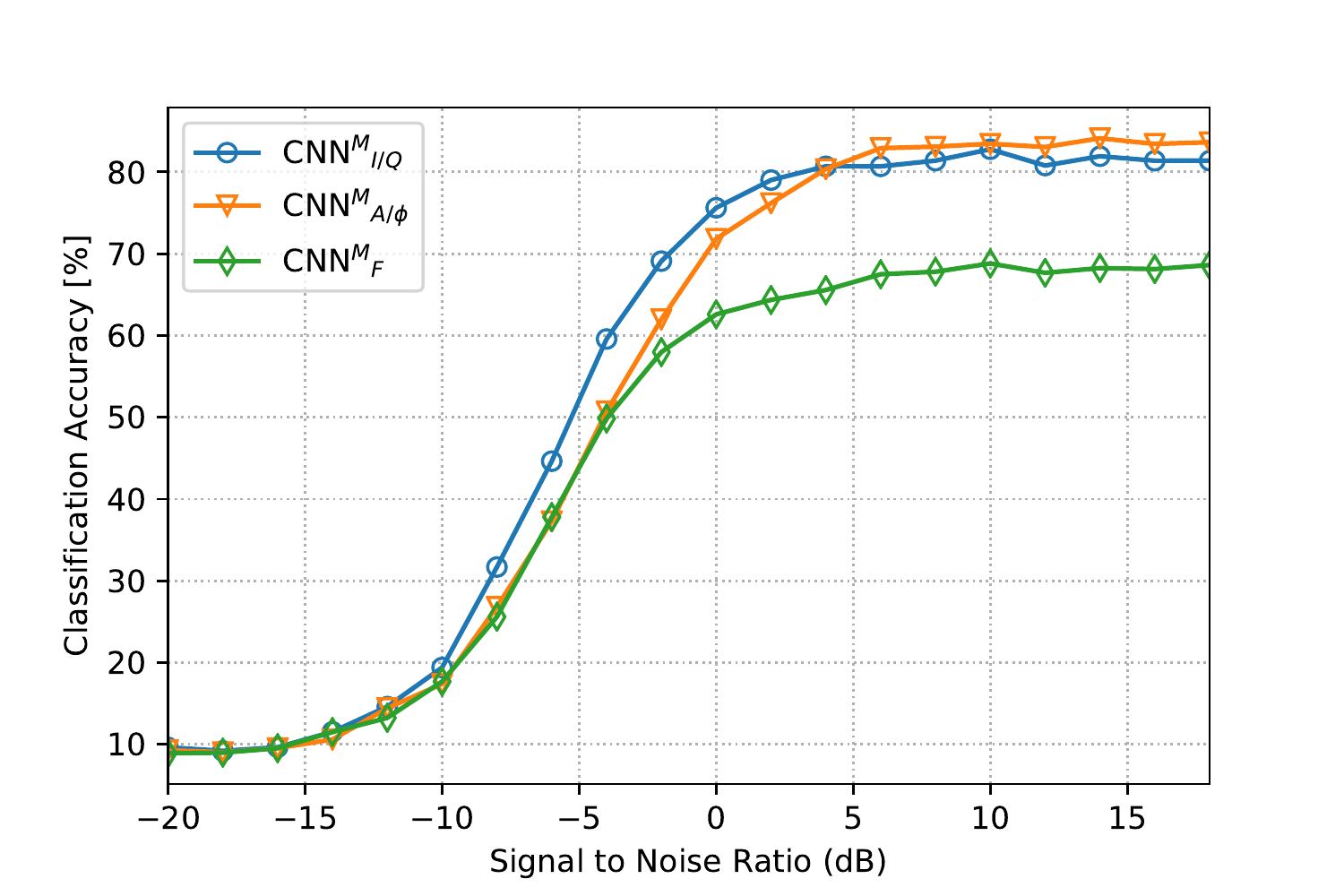}
    \caption{Performance results for modulation recognition classifiers vs. SNR}
    \label{fig:CNN_snr_mod}
\end{figure}

\textbf{Modulation recognition case.} Figure \ref{fig:CNN_snr_mod} shows that all three modulation recognition CNN models have similar performance for very low SNRs ($<-10dB$), for medium SNRs the CNN$^M_{I/Q}$ outperforms the CNN$^M_{\textbf{A}/\boldsymbol{\upphi}}$ and CNN$^M_{\mathcal{F}}$ models by 2-5dB, while for high SNR conditions ($>5dB$)  the CNN$^M_{\textbf{A}/\boldsymbol{\upphi}}$ model outperforms the CNN$^M_{I/Q}$ and CNN$^M_{\mathcal{F}}$ model with up to 2\% and 12\% accuracy improvements, respectively.
The authors in \cite{o2016convolutional} used IQ data and reported higher accuracy then the results we obtained.
We were not able to reproduce their results after various attempts on the IQ data, which may be due to the difference in the dataset (e.g. number of training examples), train/test split and hyper-parameter tuning. 
However, we noticed that the amplitude/phase representation
helped the model discriminate the modulation formats better compared to raw IQ time-series data for high SNR scenarios. We regret that results for amplitude/phase representations were not reported in \cite{o2016convolutional} too, as this may had helped improving performance.
Using the frequency spectrum data did not improve the classification accuracy compared to the IQ data. This is expected as the underlying dataset has many modulation classes, which exhibit common characteristics in the frequency domain after the channel distortion and receiver imperfection effects, particularly QPSK, 8PSK, QAM16 and QAM64. This makes the frequency spectrum a sub-optimal representations for this classification problem.

\textbf{Interference detection case.}
The IF identification models on Figure \ref{fig:CNN_snr_if} show in general better performance compared to the modulation recognition classifiers, where the CNN$^{IF}_{\mathcal{F}}$ showed best performance during all SNR scenarios. In particular, for low SNR scenarios
significant improvements can be noticed compared to the CNN$^{IF}_{\textbf{A}/\boldsymbol{\upphi}}$ and CNN$^{IF}_{I/Q}$
models with a performance gain improvement of at least 
$\sim 4dB$, and classification accuracy improvement of at least $\sim 9\%$.
The authors of \cite{schmidt2017wireless} used IQ and FFT data representations and reported similar results as our CNN$^{IF}_{I/Q}$ and CNN$^{IF}_{\mathcal{F}}$ models. However, again we noticed that the amplitude/phase representation is beneficial for discriminating signals compared to raw IQ data. But the IF identification classifier performed best on FFT data representations. This may be explained by the fact that the wireless signals from the ISM band standards (ZigBee, WiFi and Bluetooth) have more expressive features in the frequency domain as they have different frequency spectrum characteristics in terms of bandwidth and modulation/spreading method.

\begin{figure}[!t]
    \centering 
    \includegraphics[width=0.5\textwidth]{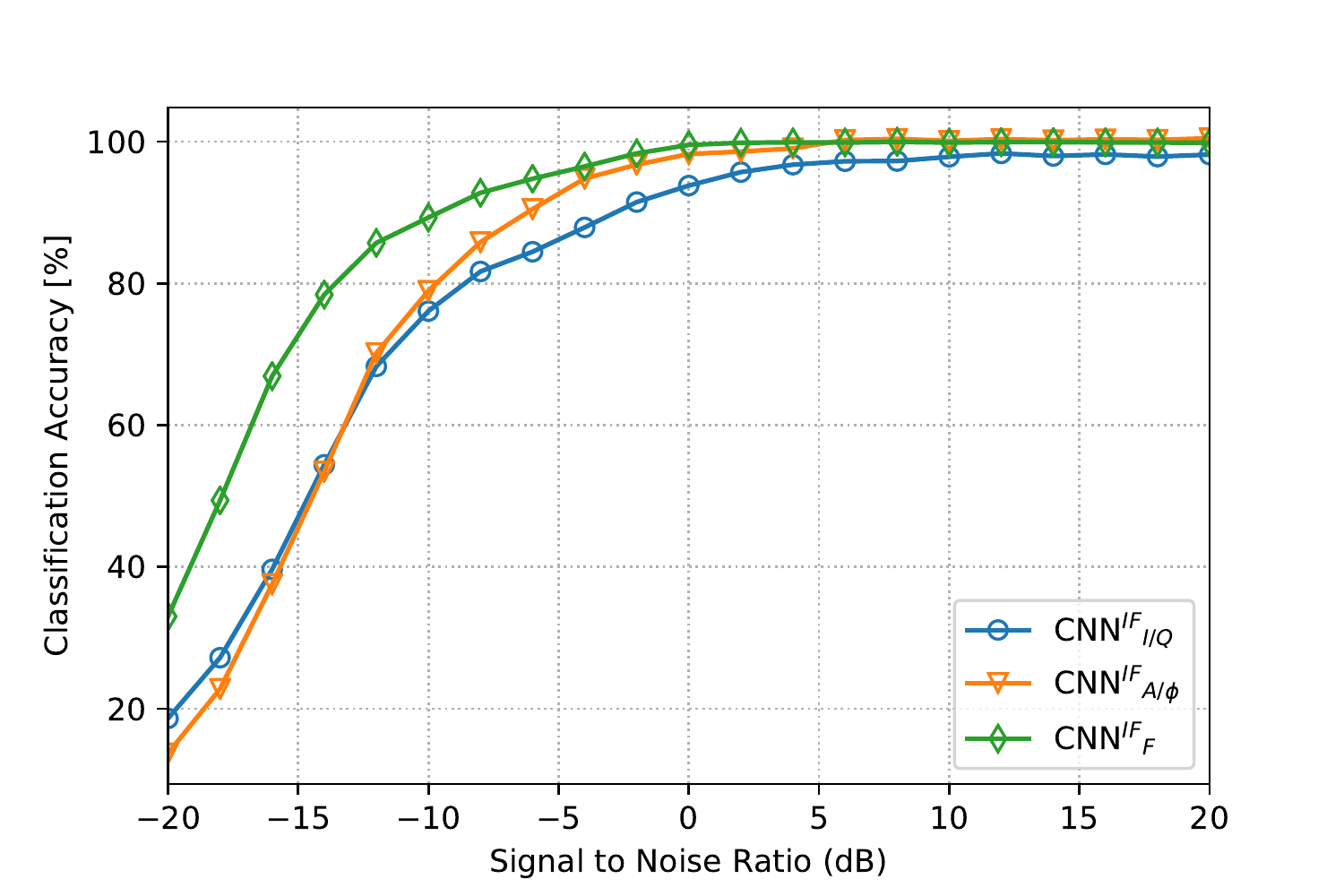}
    \caption{Performance results for interference identification classifiers vs. SNR}
    \label{fig:CNN_snr_if}
\end{figure}

\subsubsection{Takeaways}
End-to-end learning is a powerful tool for data-driven spectrum monitoring applications. It can be applied to various wireless signals to effectively detect the presence of radio emitters in a unified way without requiring design of expert features. Experiments have shown that the performance of wireless signal classifiers depends on the used data representation. This suggests that investigating several data representations is important to arrive at accurate wireless signal classifiers for a particular task.
Furthermore, the choice of data representation depends on the specifics of the problem, i.e. the considered wireless signal types for classification.
Signals within a dataset that exhibit similar characteristics in one data representation are more difficult to discriminate, which puts a higher burden on the model learning procedure. Choosing the right wireless data representation can notably increase the classification performance, for which domain knowledge about the specifics of the underlying signals targeted in the spectrum monitoring application can assist.
Additionally, the performance of the classifier can be improved by increasing the quality of the wireless signal dataset, by adding more training examples, more variation among the examples (e.g. varying channel conditions), and tuning the model hyper-parameters.

\section{Open challenges}
\label{sec:oc}
Despite the encouraging research results, a deep learning-based end-to-end learning framework for spectrum utilization optimization is still in its infancy.
In the following we discuss some of the most important challenges posed by this exciting interdisciplinary field.

\subsection{Scalable spectrum monitoring}
The first requirement for a cognitive spectrum monitoring framework is to have an \textbf{\textit{infrastructure}} that will support scalable spectrum data \textit{collection}, \textit{transfer} and \textit{storage}.
In order to obtain a detailed overview of the spectrum use, the end-devices will be required to perform distributive spectrum sensing \cite{liu2015heterogeneous} over a wide frequency range and cover the area of interest.
In order to limit the data overhead caused by huge amounts of I and Q samples that are generated by monitoring devices, the predictive models can be pushed to the end devices itself. 
Recently, \cite{rajendran2017electrosense} proposed \textit{Electrosense}, an initiative for large-scale spectrum monitoring in different regions of the world using low-cost sensors and providing the processed spectrum data as \textbf{\textit{open spectrum data}}. Access to large datasets is crucial for evaluating research advances and enabling a playground for wireless communication researchers interested to acquire a deeper knowledge of spectrum usage and to extract meaningful knowledge that can be used to design better wireless communication systems.

\subsection{Scalable spectrum learning}

The heterogeneity of technologies operating in different radio bands requires to continuously monitor multiple frequency bands making the \textit{volume} and \textit{velocity} of radio spectrum data several orders of magnitude higher compared to the typical data seen in other wireless communication systems such as wireless sensor networks (e.g. temperature, humidity reports, etc.). 
In order to handle this large volume of data and extract meaningful information over the entire spectrum, a scalable \textbf{\textit{platform}} for \textit{processing}, \textit{analysing} and \textit{learning} from \textit{big} spectrum data has to be designed and implemented \cite{zaslavsky2013sensing}, \cite{ding2014big}.
Efficient data processing and storage systems and \textbf{\textit{algorithms}} for \textit{massive} spectrum data analytics \cite{sandryhaila2014big} are needed to extract valuable information from such data and incorporate it into the spectrum decision/policy process in real-time.

\subsection{Flexible spectrum management}

One of the main communication challenges for 5G will be inter-cell and cross-technology interference. 
To support spectrum decisions and policies in such complex system, 5G networks need to support an architecture for flexible spectrum management.

\textit{Software-ization} at the radio level will be a key enabler for flexible spectrum management as it allows \textit{automation} for the collection of spectrum data, flexible \textit{control} and \textit{reconfiguration} of cognitive radio elements and parameters.
There are several individual works that focused on this issue. Some initiatives for embedded devices are WiSCoP
\cite{kazaz2016wiscop}, Atomix \cite{bansal2015atomix} and \cite{kazaz2017hardware}.
Recently, there is also a  growing interest in academia and industry to apply Software Defined Networking (SDN) and Network Function Virtualization (NFV) to wireless networks \cite{zaidi2017will}.
Initiatives such as SoftAir \cite{akyildiz2015softair}, Cloud RAN \cite{checko2015cloud}, OpenRadio \cite{bansal2012openradio} and several others are still at the conceptual or prototype level.
To bring flexible spectrum management strategies into realization and the commercial perspective a great deal of standardization efforts is still required.

\subsection{Spectrum privacy}

The introduction of intelligent wireless systems raises several privacy issues. The spectrum will be monitored via heterogeneous radios including WSNs, RFIDs, cellular phones and others, which may lead to misuse of the applications and cause severe privacy-related threats. Therefore, privacy is required at the spectrum data collection level. As spectrum data may be shared along the way, privacy has to be maintained also at data sharing levels. Thus, \textbf{\textit{data anonymization}}, restricted \textbf{\textit{data access}}, proper \textbf{\textit{authentication}} and strict control of intelligent radio users is required. 


%
\section{Conclusion}
\label{sec:concl}
This paper presents a comprehensive and systematic introduction to \textit{end-to-end} learning from spectrum data - a deep learning based unified approach for realizing various wireless signal identification tasks, which are the main building blocks of spectrum monitoring systems.  
The approach develops around the systematic application of 
deep learning techniques to obtain accurate wireless signal classifiers in an end-to-end learning pipeline.
In particular, convolutional neural networks (CNNs) lend themselves well to this setting, because they consist of many layers of processing units capable to (i) automatically extract non-linear and more abstract wireless signal features that are invariant to local spectral and temporal variations, and (ii) train wireless signal classifiers that can outperform traditional approaches.


With the aim to raise awareness of the potential of this emerging interdisciplinary research area, first, 
machine learning, deep learning and CNNs were briefly introduced and a reference model for their 
application for spectrum monitoring scenarios was proposed.
Then, a framework for end-to-end learning from spectrum data was presented. In particular, wireless data collection, the design of wireless signal features and classifiers suitable for several wireless signal identification tasks are elaborated. Three common wireless signal representations were defined, the raw IQ temporal wireless signal, the time domain amplitude and phase information data, and the spectral magnitude representation.
The presented methodology was validated on two active wireless signal identification research problems: (i) \textit{modulation recognition} crucial for dynamic spectrum access applications and (ii) \textit{wireless interference identification} essential for effective interference mitigation strategies in unlicensed bands. 
Experiments have shown that CNNs are promising feature learning and function approximation techniques, well-suited for different wireless signal classification problems. Furthermore, 
the presented results indicated that for the wireless communication domain investigating different wireless data representations is important to determine the right representation that exhibits discriminative characteristics for the signals that need to be classified.
Specifically, in the modulation recognition case study for medium-high SNR the CNN model trained on amplitude/phase representations
outperformed the other two models with a $~2\%$ and $~10\%$ performance improvement, while for low SNR conditions the model trained on IQ data representations showed best performance.
For the task of detecting interference, the model trained on FFT data outperformed amplitude/phase and IQ data representation models by up to $20\%$ for low SNR conditions, while for medium-high SNR up to $5\%$ classification accuracy improvements.

These results demonstrate the importance of both choosing the correct data representation and machine learning approach, both of which are systematically introduced in this paper. By following the proposed methodology, deeper insights can be obtained regarding the optimality of data representations for different research domains. As such, we envisage this paper to empower and guide machine learning/signal processing practitioners and wireless engineers to design new innovative research applications of end-to-end learning from spectrum data that address issues related to cross-technology coexistence, inefficient spectrum utilization and regulation. 

\section{Acknowledgements}
We are grateful to Prof. Gerard J.M. Janssen for his insightful
comments, and Malte Schmidt et al. \cite{schmidt2017wireless} for sharing the "Wireless interference" dataset. 
This work was partly supported by the EU H2020 eWINE project under grant agreement number 688116, SBO SAMURAI project and the AWS Educate/GitHub Student Developer Pack.

\ifCLASSOPTIONcaptionsoff
  \newpage
\fi




\bibliographystyle{IEEEtran}
\bibliography{biblio}

\end{document}